\documentstyle[12pt,aasms4]{article}

\begin{document}
\begin{titlepage}
\title{ The Globular Cluster System in the Inner Region of M87\altaffilmark{1}    }

\author{ Arunav Kundu\altaffilmark{2}, Bradley C. 
Whitmore, William B. Sparks 
and F. Duccio Macchetto }
\affil{ Space Telescope Science Institute, 3700 San Martin Drive, Baltimore, MD 21218 }
\affil{  Electronic Mail: akundu whitmore sparks macchetto, @stsci.edu}
\author{ Stephen E. Zepf } 
\affil{ Department of Astronomy, Yale University, P.O.Box 208101, New Haven,
CT 06520-8101 }
\affil{  Electronic Mail: zepf@astro.yale.edu }

\author{ Keith M. Ashman } 
\affil{ Department of Physics and Astronomy, University of Kansas, Lawrence, KS
 66045 }
\affil{  Electronic Mail: ashman@kusmos.phsx.ukans.edu }
\altaffiltext{1}{\normalsize Based on observations with the 
NASA/ESA {\it Hubble Space Telescope}, obtained at the Space Telescope Science 
Institute, which is operated by the Association of Universities for Research 
in Astronomy,Inc., under NASA contract NAS5-26555 }
\altaffiltext{2}{ Astronomy Department, University of Maryland,  College Park, MD 20742 }
 \begin{abstract}
	
	 1057 globular cluster candidates have been identified in a WFPC2 image 
of the inner region of M87. The Globular Cluster Luminosity Function (GCLF) can be well fit by a Gaussian profile with a mean value of
 m$_V$$^0$=23.67$\pm$0.07 mag and $\sigma$=1.39$\pm$0.06 mag (compared to  m$_V$$^0$=23.74
and  $\sigma$=1.44 from an earlier study using the same data  by Whitmore {\it et al.} 1995).  The GCLF in five radial 
bins is found to be statistically the same at all points, showing
 no clear evidence 
of dynamical destruction processes based on the luminosity function (LF), in 
contradiction to the claim by Gnedin (1997). Similarly, there
is  no obvious correlation between the half light radius of the clusters and the
 galactocentric distance. The core radius of the globular cluster density distribution  is R$_c$=56$''$, considerably larger than the core of  the stellar component (R$_c$=6.8$''$). The  mean color 
of the cluster candidates is V-I=1.09 mag which corresponds to an average metallicity of Fe/H = -0.74 dex. The color distribution  is bimodal everywhere, with a
blue peak at V-I=0.95 mag and a red peak at V-I=1.20 mag.    The  red population is only 0.1 magnitude bluer than the
underlying galaxy, indicating that these clusters 
 formed late in the metal enrichment history
 of the galaxy and were  possibly created in a burst of star/cluster formation 3-6 Gyr after the blue population. We also find that both the red and the blue cluster distributions have a more elliptical shape (Hubble type E3.5) than the nearly spherical galaxy.  The average half light radius of the clusters is $\approx$2.5 pc which is comparable to the 3 pc average effective radius of the
 Milky Way clusters, though the red candidates are $\approx$20\% smaller than the blue ones.

\end{abstract}

\end{titlepage}
\section{Introduction}
	The numerous point sources that appear around 
 M87 (NGC 4486)  were first identified as its  globular cluster system 
by Baum (1955). Since then, the globular cluster system, hereafter GCS, of
this giant elliptical galaxy near the core of the Virgo cluster has been 
 one of the most carefully studied systems and has led the way in the study 
of globular clusters in external galaxies. Racine (1968a, 1968b) first investigated
 the magnitude and color distribution of the clusters using photographic plates 
and reported that they are on average bluer than the galaxy
 background. Strom {\it et al.} (1981) corroborated this result and also suggested
that the mean color of the clusters shows a radial trend, with the  more distant clusters being bluer than ones that are closer to the nucleus. 
More recent studies by Lee \& Geisler (1993), Cohen {\it et al.} (1998) and Nielsen {\it et al.} (1998) have confirmed the radial variation in color. 
The effective radius of the GCS was reported to be  much larger than that of the underlying galaxy by Harris \&
Racine (1979), and has since been verified by other groups
 (Harris 1986; Lauer \& Kormendy 1986; Grillmair, Pritchet \& van den Bergh 1986; McLaughlin 1995). More recently the GCS of M87 
has  been   used   to measure the Hubble constant by
exploiting the apparent constancy of the globular cluster luminosity function (GCLF).
 Whitmore {\it et al.} (1995) studied the 
globular clusters in the core region of the galaxy using the WFPC2 camera onboard the HST and  were able to detect clusters that were two magnitudes
fainter than the turnover luminosity of the GCLF for the first time, hence providing a more secure measurement of the turnover. The superior angular resolution of the 
HST enabled them to distinguish globular clusters from the galactic background
to a limiting magnitude of V=26 mag.  This dataset provides the 
best determined GCLF of {\it any} galaxy currently available, better even than 
for the Milky Way or M31 due to the much larger number of clusters.

 In this paper, we follow up on the work of  Whitmore {\it et al.} (1995), hereafter referred to as Paper 1, and study the  properties of the
 globular clusters in
 the inner region of M87 in the context of the formation and evolution of 
the GCS.

 In Paper 1  we found that the color distribution of the clusters is bimodal, 
which was confirmed by the observations of Elson \& 
Santiago (1996a, 1996b). This is consistent with the merger scenario of 
globular cluster formation  proposed by Ashman \&
 Zepf (1992), who suggested that globular clusters can be formed by the
interaction or merger of galaxies. These clusters would be more metal
rich, and hence redder, than the older population associated with the 
interacting galaxies (see Ashman \& Zepf 1997 for a detailed discussion). Developing 
upon the merger scenario Whitmore {\it et al.} (1997) have modelled the evolution of the color distribution of globular clusters with time, and they suggest an age of 9$\pm$5 Gyr for the population of red clusters in M87.
 In this paper we study the two populations
 and  attempt to resolve questions about their formation and evolutionary histories.

It has been suggested that the universality of the turnover luminosity of the 
GCLF is caused by destruction mechanisms that conspire to 
preferentially destruct clusters outside a  narrow range of luminosities, 
and hence masses,  near the turnover (Fall \& Rees 1977).
 The major players are dynamical friction that preferentially destroys 
the high mass clusters, and  tidal shocking and evaporation which act more efficiently on low density,
low mass clusters (Murali \& Weinberg 1997a, 1997b; Ostriker \& Gnedin 1997;
 Gnedin \& Ostriker 1997 and references therein). 
Theory predicts the range of their maximum efficiency in the region of 2-8 kpc of the 
galactic center,  depending on the  model.
 If this is true then the luminosity distribution should vary most strongly 
in this region, and it raises concerns about the 
suitability of using an universal luminosity function to describe the distribution 
of cluster luminosities. However, previous studies (Grillmair {\it et al.}
 1986; Lauer \& Kormendy 1986; McLaughlin, Harris \& Hanes 1994) have found no evidence for variation of 
the GCLF with radial distance from the center of the galaxy. The only
claim to the contrary  is by   Gnedin 
(1997) who reports that the clusters in the inner region are brighter
 than those in the outer region by 0.26 mag. In order to 
examine the importance of destruction 
mechanisms in modifying the globular cluster system we study the variation of 
the GCLF and cluster density with galactocentric distance. We also measure the sizes of the clusters  to determine if there is a radial variation in the size distributions 
due to dynamical evolution of the clusters.
 
\section{Observations and Reductions}

	We observed the inner region of M87 with the Wide Field and Planetary
 Camera 2 (henceforth WF=Wide Field Camera and PC=Planetary Camera) onboard 
the HST on February 4, 1995. Four 600 secs exposures were taken with 
the broad band filters F555W  and F814W. These four separate exposures provide 
excellent cosmic ray removal, 
which was performed using the STSDAS task GCOMBINE. This was followed by the IRAF task 
 COSMICRAYS to remove hot pixels. Fig 1 shows the mosaiced V (F555W) image 
from the four WFPC2 chips.  Numerous cluster candidates can easily be
 identified.  Fig 2 is the residual V image of the region near the center of the
 galaxy after a median filtered image has been subtracted.  It
 demonstrates the superior
 ability  of the HST to resolve point sources in the presence of a strong
 background. The optical jet and several cluster candidates can easily be seen 
right into the core. 

\subsection{Object Identification}

	The initial detection of candidate clusters was carried out using the 
 DAOFIND routine in DAOPHOT (Stetson 1987) using the V + I image. This list  was
 then refined by manual inspection. The criteria that we used to manually 
identify clusters 
were: 1) the objects must be point-like, 2) they must
be present in the V, I and V+I image, 3) at least 3 adjacent pixels must
be above the background level, 4) the objects must have a reasonable shape
(e.g. all the bright pixels should not be along a column or a row). At 
this stage, only the sources 
that obviously appeared to be extended and likely to be background galaxies 
were eliminated. During this process $\approx$10\% of the objects detected by
 DAOFIND were rejected and  another $\approx$15\% added to the list.

	Another object list was produced using a completely objective method of 
identification similar to the one used by Whitmore {\it et al.} (1997). One of the shortcomings of  DAOFIND is that it uses
 a constant value of $\sigma$ for 
the background noise, which leads to spurious detection of objects in 
regions of high noise (corresponding to regions of high surface brightness). To
 overcome this problem, we developed a method of identifying sources
using the local S/N ratio. Initially, potential objects in the V+I image
were identified by using a very low cutoff (2$\sigma$) in DAOFIND.
This step typically returned a few thousand sources in each of the chips. 
Aperture photometry was then performed on each of these sources with radii of
 0.5 pixels and 3 pixels and a sky annulus between 5 and 8 pixels.
 The standard deviation from the median of the sky value for 
each source was taken to be a measure of  the background
noise at that point. At this stage, any object where the flux within 0.5 pixels was
at least 3 times greater than the standard deviation of the sky pixels, or was at 
least 20 counts above the 
background, was identified as a candidate cluster.  Aperture photometry was then
 performed on this list of objects in  the F555W 
 and F814W images separately. Only objects that had  a S/N $>$ 1.5 in both the 
images were considered bona fide detections.

From our aperture correction tests (see $\S$ 2.2) we found that for a typical
 cluster in the PC, the flux within 3
pixels is  $\approx$7.5 times the flux within 0.5 pixels and the ratio of fluxes is 
$\approx$5.5 for the WF. Allowing for variation in the sizes of the clusters we concluded that $\frac{Counts_{3pix}}{Counts_{0.5pix}}$ $<$ 12 for the PC and  
 $\frac{Counts_{3pix}}{Counts_{0.5pix}}$ $<$ 8 for the WF is a good discriminant to weed 
out background galaxies (This is similar to the concentration criteria used in 
Paper 1). Additionally, a lower limit cutoff of  $\frac{Counts_{3pix}}{Counts_{0.5pix}}$ $>$ 1.5
 is useful in rejecting
chip defects or hot pixels that might be present after the 
initial reduction. Both our final objective list and the DAOFIND/manual object
 lists were  passed through these filters.
There was excellent agreement between the sources found by the two methods. The
objective method identified 1057 sources in all 4 chips, while 1012 sources were
 found by the manual method. Of these, 989 sources were identical in the two lists. 
The cluster candidates that did not match  were either near the detection limit 
and/or had large photometric errors.  A comparison of the photometry based on
 the manual and objective lists reveal no significant effect of different
selection criteria, hence we use the objective list for further analysis.

	The V and I images were also cross correlated to search for spatial shifts
between the two sets. The PC images were found to have offsets less than 0.07
pixels while the WF chips had offsets within 0.035 pixel. Due
to the excellent spatial registration of both sets of images, identical pixel
positions, determined from the V image, are used for the rest of the analysis.

 \subsection{Aperture Photometry}

	Close inspection of the radial profiles of the cluster candidates show
that  they are statistically broader than true point sources. We
 use this to study the sizes  of the clusters in $\S$ 3.7.
 However, this  also means that the aperture
corrections derived for point sources would be slightly small if applied to 
the clusters in M87. In order to quantify this difference we computed the flux within
 apertures of various sizes on a sample of  bright clusters in all the chips and compared
these  with profiles for stars in the calibration field NGC 5139
($\omega$ Cen) and the standard star GRW+70D5824. The average cluster
is found to be 0.09 mag brighter in both V and I when compared to numbers
reported in Paper 1 using point source corrections (Holtzman {\it et al.} 1995a).
{\it This 0.09 mag difference, and the more objectively determined candidate list,
 are the primary changes between this paper and Paper 1.}
	
	The photometric zeropoints are adopted from Whitmore {\it et al.} (1997)
who derived the values by comparing HST and ground based photometry for 6 galaxies. The average of 
the 6 galaxies yields  a zeropoint of 22.573 mag  to convert from F555W to 
Johnson V and 21.709 mag to convert
from F814W to Cousins I. These numbers agree closely with the Holtzman {\it et al.} 
(1995b) zeropoints of 22.56 mag in V and 21.69 mag in I. 

 In Paper 1 we used Burstein and Heiles's (1984) value 
of A$_B$=0.09$\pm$0.06 mag and reddening curves from Mathis (1990) to derive an 
extinction of A$_V$=0.067$\pm$0.04 mag and A$_I$=0.032$\pm$0.02 mag.  Sandage \& 
Tammann (1996), commenting on Paper 1, argue that the extinction in the 
direction of M87 should be 0.00 mag. We continue to use the above values of the 
extinction to maintain continuity with Paper 1 since there is no compelling 
reason to assume that there is no reddening in the direction of M87.

  The zeropoints and aperture corrections  used to perform aperture photometry
 using the PHOT task in the DAOPHOTX package within IRAF  are listed 
in Table 1. A correction of 0.1 magnitude has been added to normalize
the measurements made within a radius of 0.5$''$ to infinite radius
 following Holtzman {\it et al.} (1995b). Since the difference between the aperture corrections in the three WF chips is smaller than the rms variation within a 
chip, we have used the
 same correction for all the WF chips to determine the brightness of the 
candidate globular clusters. An aperture radius of 2 pixels  for the WFs, and 3 pixels for the PC is chosen so that the percentage of encircled energy is approximately the same on  both chips (Note: an aperture radius of 3 pixels was used for both 
the PC and the WF in Paper 1). The sky annulus is defined to be between 5-8 
pixels, fairly close to the
source, because the large galaxy background gradient introduces errors in
the determination of the background when an aperture with a larger radius is 
used.	Table 2 lists the magnitudes and positions of the  100 clusters closest
to the nucleus. 
	
\subsection{Completeness Corrections}

	For statistical studies of  globular cluster systems it is
 necessary to know the detection limit of clusters in the field. To quantify
this effect  completeness tests were performed on the images. 
 First, clusters with high S/N were extracted from regions
of low background noise to be used as model objects for the tests. A random number 
routine was used
to add objects of known magnitude and V-I colors between 0.9 mag and 1.3 mag. In each pass one hundred 
objects were added per chip and the objective algorithm
described above was then used to detect candidate
clusters in the chips. In all, approximately 10000 simulated objects 
were added to the two sets of images in V and I. The completeness limits 
determined from these tests are plotted in Fig 3. It can 
be seen that the detection limit migrates
toward  progressively fainter magnitudes with increasing distance from the 
center of the galaxy, as is to be expected from the background light 
distribution. In the innermost region of the PC (0-10$''$) the 50\% completeness limit is at V=24.3 mag while in the region between 20-30$''$, it is at V=25.3
mag. The  detection limit in the WF chips varies from V=24.1 mag to V=25.7 mag 
while the average  50\% threshold for the entire field (WF+PC) is at V=25.5 mag.  

	At first glance it might seem surprising that the detection threshold in
 the innermost bin of the WF (6-30$''$) is brighter than the threshold in the PC.
 However, the galaxy background at comparable radii is larger for the WF than the 
PC due to the larger linear size of the WF pixels. Since the nucleus of the galaxy
 is near the apex of the WFPC2 the background counts per pixel due to the 
galaxy is larger in the inner regions of the WF than everywhere but  very 
near the center of the  PC. The Poisson noise in the background is the limiting factor in 
object identification, 
hence the lower threshold.

\subsection{Contamination Corrections}

	One of the biggest advantages of the high resolving power of the HST is that 
background galaxies can be identified and rejected fairly easily. However, the
background galaxy count increases sharply near the limit of our completeness
 threshold and a few are probably still lurking in our object lists 
masquerading as clusters. Furthermore, the lists may be contaminated by 2-4
stars in each of the chips. In order to quantify this  correction, we used the automated routine  to identify sources  in two images,
 one each in F606W and F814W, from a Medium Deep Survey  field (Griffiths {\it et al.} 1994) located about 
9.3 degrees from the 
program galaxy. We then performed aperture photometry on the objects that
passed our normal filters for  cluster candidates, taking into
 account  an offset of 0.29 magnitudes for the photometric transformation 
from F606W to F555W  (derived using the SYNPHOT package). Fig 4 shows the 
luminosity function for  these objects. There are only 
about 20 contaminating objects with V$<$25 mag within the WFPC2 field of view 
but the number  increases sharply near V=25.5 mag.

\subsection{Surface Photometry}

	One of the  main aims of this paper is to study the spatial variations
 of various properties of the globular clusters in M87 and to compare them to the underlying 
galaxy. A previous study by Boroson, Thompson \& Shectman (1983) found that V-I$\approx$1.6 mag
in the inner regions of the galaxy and decreases with radius. On the 
other hand, Zeilinger, Moller \& Stiavelli
(1993), employing more modern CCD techniques, reported that the V-I color 
distribution of the galaxy is nearly constant at 1.3 mag in the inner region of 
the galaxy. To resolve this issue we  derived the surface 
profiles of the galaxy from our image. First,  the images from the 4 CCDs
 were mosaiced using the STSDAS task WMOSAIC. Then, after visual inspection, 
a  mask was used to blot out the optical jet and the blank region outside the PC. The STSDAS
 tasks ELLIPSE and BMODEL were  used to model the smooth galaxy background and 
fit  circular profiles with a fixed center. Sparks, Laing \& Jenkins (1988) report 0.08$<$10(1-$\frac{a}{b}$)$<$0.79 in the inner 1 arcmin of 
the galaxy. Thus, the ratio of the axes $\frac{a}{b}$
is between 0.92 and 0.99 and the assumption of circular isophotes does not 
introduce a significant source of error.

The small field of view of the WFPC2 presents some problems because the nuclear
 region of M87 completely fills
the aperture, making accurate sky background measurements difficult.
	The background counts were calculated from the average 
brightness of  the MDS field used for background object detection. After applying a
photometric correction  to the  sky intensity  found in the F606W image  in order to  convert to F555W  the derived background level in the V image is 
9.2 counts per pixel. The
corresponding correction for the  F814W image is  6.0 counts per pixel. While this method of calculating 
the background is not very secure, we found that subtracting background counts of this order 
 affected the color profile by less than 0.05 mag (vis-a-vis no background correction) in the inner 15$''$ 
of the galaxy where it is brightest. On comparing our V and I band profiles with that  of 
 Zeilinger {\it et al.} (1993) we find that they match up very well in both the 
bandpasses. The  V-I color of the galaxy
 in this region is constant at 1.3 mag  and is consistent with the photometry of
 Zeilinger {\it et al.} (1993). On the other hand our V-band profile  matches the  Boroson {\it et al.} (1983) data quite well, but our I band profile is $\approx$0.2 mag dimmer than
 theirs. This difference may be because Boroson {\it et al.} use a Gunn i filter (Wade {\it et al.} 1979) which has an effective wavelength 
$\lambda_{i}$=8200A,  while our I magnitudes are in the Cousins  system
 ($\lambda_{I}$=8000A). Therefore,  we use Zeilinger et als (1993) photometry for further analysis and interpretation.

\section{Results and Discussion }

\subsection{ The Luminosity Function}

	The most significant result reported in Paper 1 was that the GCLF can
 be measured roughly 2 magnitudes deeper than the turnover luminosity. We have re-evaluated
 these calculations to see how the numbers change due to the new detection 
routines and aperture corrections.

	In Fig 5 we plot the luminosity function for the object list used 
in Paper 1 and the objectively identified candidate 
list from the current paper. The  new aperture,  completeness, and background 
corrections from Table 1, Fig 3, and Fig 4 respectively, have been applied to the
 GCLF in the top panel, the objectively identified clusters, while the bottom
panel is a plot of the luminosity function calculated in Paper 1.
 The turnover magnitude in the V band 
 for the objective set is  m${_V}{^0}$=23.67$\pm$0.06 mag for the best fitting Gaussian compared to  
m${_V}{^0}$=23.74$\pm$0.06 mag in Paper 1. We use the objectively 
identified list as our best estimate, hence the turnover is 0.07 magnitudes
brighter than that reported in the earlier analysis, primarily due to the new aperture corrections. A similar analysis of the luminosity function in the I
 band gives  m${_I}{^0}$=22.55$\pm$0.06 mag, as compared to the Paper 1 value of 
 m${_I}{^0}$=22.67 mag. 

\subsection{Spatial Variation of the Luminosity Distribution }

	The fact that the GCLF of luminous elliptical 
galaxies appears to be nearly universal (Harris 1991; Jacoby {\it et al.} 1992;
 Whitmore 1996; Ashman \& Zepf 1997) is a rather remarkable result which  
implies that, for any  reasonable range of M/L
 ratio,  the underlying mass distribution of globular clusters is similar.
 It is rather amazing that the the GCLF of a large number of galaxies
of various shapes and sizes have similar GCLFs in spite of the  destruction
 mechanisms that may be constantly acting upon the clusters with a relative efficiency
that depends strongly on the shape and/or size of the galaxy. In particular,
theoretical models predict that these processes should have a 
measurable effect in the inner regions of a galaxy, with
dynamical friction and tidal shocking being suggested as the most 
important mechanisms (Aguilar, Hut \& Ostriker 1988; Murray \& Lin 1992). While dynamical friction operates more efficiently on high mass
 clusters and causes them to spiral towards the nucleus of the galaxy,  tidal 
shocking selectively destroys the low density clusters.  There is
 the added possibility that the clusters are destroyed or captured by a massive central
 object e.g. a blackhole (McLaughlin 1995). Even though all of
these mechanisms are predicted to be most efficient from anywhere within the 
inner 2 kpc to the inner 8 kpc depending on the  model, most  observations of the
 core region of M87 (Grillmair {\it et al.} 1986; Lauer \& Kormendy 1986; McLaughlin,
 Harris \& Hanes 1994; Harris, Harris \& McLaughlin 1998)  have 
revealed no evidence of a radial variation of the GCLF. The only claim to the
contrary has been made by Gnedin (1997), who suggests that the clusters in the 
inner regions of M87 are systematically brighter than those in the outer
regions by 0.26 magnitudes, which he attributes to the preferred destruction
 of low mass clusters in the stronger tidal field of the inner regions of the galaxy. However, ground based data may be strongly affected by  selection effects
which can lead to spurious results e.g. fainter (low mass) clusters are harder 
to detect near the center of the galaxy where the background light is the
 strongest and it can point towards an apparent radial variation in intensity if
this effect is not accurately compensated for. Our data goes much deeper than previous observations, with  higher completeness fractions near the center of 
M87, and hence gives 
us a better chance of identifying any radial variations in the LF 
 and addressing the question of the universality of the GCLF.

	Fig 6 is a plot of the radial distribution of cluster 
magnitudes for all objects with photometric errors less than 0.3 magnitudes. A 
least squares fit straight line shows a weak correlation between the brightness
of the clusters and the galactocentric radius, but this is almost 
certainly an artifact of the fainter completeness limits at larger radii.
	In order to make a more meaningful test of the variation in the LF with 
distance, we divided the sample into 5, roughly equal, radial bins and corrected each of distributions  for completeness and contamination. Both the raw and
the corrected luminosity distribution for each of the bins, in both the V and
the I bands, is plotted in Fig 7.  Close inspection of the corrected profiles indicates that the GCLF is  remarkably similar at all radial distances. In order
 to quantify this observation we fit Gaussian profiles  to the corrected
 luminosity distributions.  The peak 
magnitude and $\sigma$ (the dispersion) for the best fitting Gaussian is plotted in Fig 8.  Although the four inner bins of the V band 
data do seem to show  that the turnover magnitude is brighter in the 
inner bins, the outermost bin, for which the turnover luminosity is most securely determined, does not conform to this trend. The turnover luminosity 
of the innermost bin is less than 1$\sigma$ brighter than the mean, while the 
peak of the bins at 53.5$''$ and 66.6$''$ are $\sim$1$\sigma$ dimmer than the mean. The I band luminosities show no significant radial trends whatsoever. 
Therefore it appears that both the peak
of the distribution and the half-width  are   constant to within the 
uncertainties, with no obvious radial dependance that can be attributed to a 
particular destruction mechanism.   At this point one could argue that the ad hoc
choice of a Gaussian has no physical basis, and that a variation
 in the shape of the luminosity function may be hidden under the parameters
of the Gaussian fit. In order to address this issue 
we plot the cumulative profile of the corrected distributions in 5 radial 
bins in Fig 9. We have considered only clusters brighter than 24 mag in V, and
 23 mag in I, where the completeness and background correction are minimal.
Since the normalized  profiles in neither the V, nor the I band show a
 consistent, distance dependant variation, we are led to conclude that we see no
 believable evidence of the effect of destruction mechanisms in the luminosity
function of M87. K-S tests, performed  on the unbinned data in the bright portion of the distributions which are unaffected by varying completeness limits, confirm that
 all the distributions are statistically identical (with typical confidence limits of 70\%  that the distributions are related).   Our conclusion does not agree with Gnedin (1997), who claims to see evidence of cluster destruction based on his interpretation of the
 McLaughlin, Harris \& Hanes (1994) data. Gnedin finds that the turnover magnitude of the GCLF in his inner region (1.21$'$$<$R$<$3.64$'$) at 
V=23.13 mag, is 0.24 mag brighter than the turnover magnitude of his outer
 region (3.64$'$$<$R$<$6.82$'$) at V=23.37 mag, which he interprets as
evidence of tidal shocks in the inner region. If this is a real physical effect 
then we expect  our sample of clusters, which have an
average radial distance R$\approx$0.83$'$, to have a turnover magnitude brighter
than Gnedin's inner clusters. However, the overall turnover luminosity
of our sample at V=23.67 mag as well as the turnover luminosities of each of 
the radial bins (Fig 8) are dimmer than even Gnedin's outer sample and are clearly 
at odds with his data. Though unlikely, this discrepancy could possibly be
due to a zerpoint error in either our photometry or that of the McLaughlin, 
Harris \& Hanes (1994) data on which Gnedin bases his analysis. However, as Harris 
{\it et al.} (1998) find no zeropoint offset between their observations and the 
McLaughlin at al. (1994) data, and the turnover luminosity of the Harris et 
al. (1998) observations matches our's (they estimate m$^v$$_0$=23.7 mag) we conclude that
there the offset is unlikely to be due to zeropoint errors.     We believe that the apparent brightening observed by 
Gnedin is probably due to undercompensation of completeness corrections in the inner
regions where the dimmer clusters are harder to detect against the strong 
galaxy background. 
	Another interesting result that we see in Fig 8 is that the scatter
in the turnover magnitude and $\sigma$ is marginally smaller in the I band than 
in the V band ($m{_V}{^0}$~=~23.71$\pm$0.11 mag; $m{_I}{^0}$~=~23.67$\pm$0.06 mag:  $\sigma{_V}$~=~1.37$\pm$0.16 mag; $\sigma{_I}$~=~1.41$\pm$0.11 mag).  This agrees with  the suggestion of Ashman, Conti \& 
Zepf (1995) that the luminosity function in the I band is less affected  by
 variations in the metallicities of the clusters and may be a better choice 
for distance measurements.

\subsection{ The  Color Distribution }

As discussed in Paper 1, the color distribution of the globular clusters
 in the inner regions of  M87 is bimodal. The mean color of all the clusters was estimated 
to be V-I~=~1.10$\pm$0.01~mag, with the blue peak at V-I~=~0.95~mag and the red
 peak at V-I~=~1.20~mag. The color distribution of 
clusters having
 photometric error less than 0.2 magnitudes in V-I, as derived in Paper 1, is
compared  with the current list in Fig 10. 

The mean color of the clusters in the objectively identified list is 
V-I~=~1.09$\pm$0.01 ~mag. A simultaneous fit of two Gaussians to 
the color distribution of the data shows that  the blue and 
the red peaks are at V-I~=~0.95$\pm$0.02 mag and V-I~=~1.20$\pm$0.02 ~mag
respectively, which, within the uncertainties, is identical to the peak values of
 0.97$\pm$0.02 ~mag and 1.21$\pm$0.02 ~mag found
by fitting Gaussians to the Paper 1 list. So even though the luminosity function of the Paper 1 list and the objectively identified list are slightly different due to the new aperture corrections, the 
color distribution are nearly identical.

	We  also use the KMM mixture modelling algorithm, described in Ashman, Bird
\& Zepf (1994), to independently  test for bimodality, and to partition the objectively identified candidate list into
 sub-populations. Since KMM is sensitive to outliers in the dataset, only 
 the candidate objects that are within the range 0.6$<$V-I$<$1.6 are 
considered. This reduces the number of objects to 997. We find that the 
distribution can be divided into 2 sub-populations. Though Lee \& Geisler (1993)
suggest that the distribution may be trimodal we find no reasonable
partition with 3 groups that supports this claim.  In the case of a homoscedastic
partition (2 populations forced to have equal variances) there are 428 blue
objects with mean V-I=0.963 ~mag and 569 red objects with V-I=1.201 ~mag
   and  a common dispersion of 0.134 mag. The threshold color dividing the two 
populations is V-I=1.064 which is slightly smaller than the value derived from
 fitting Gaussians. For a heteroscedastic partition (red and blue populations
allowed different variances) we obtain  336 blue candidates with a mean 
V-I=0.935 ~mag and 0.123 mag dispersion, and 661 red candidates with V-I=1.18
 mag and 0.141 mag dispersion. We shall adopt V-I=0.95 mag as the peak of the blue clusters, V-I=1.20 mag as the peak of the red clusters, and V-I=1.09 as
the mean of the entire distribution, for the rest of the paper. From hereon 
we define clusters that have  V-I$<$1.09 mag as blue clusters and those with 
V-I$>$1.09 as red clusters.

	 The  relationship  between the broad band colors and metallicities of globular clusters is known to be roughly linear, and the most commonly used
expression,  derived by Couture, Harris \& Allwright (1990), using the Milky
Way globular clusters, is V-I = 0.198[Fe/H] + 1.207. However, we found in Kundu
\& Whitmore (1998), that the slope of the equation changes significantly due
to the choice of the independent variable used to derive this relationship and
that the above equation probably overestimates the metallicity of metal rich
clusters.  Therefore we used the following relationship derived (using Milky Way clusters)
 in Kundu \& Whitmore (1998) to convert broad band colors to metallicities:
\begin{equation}  
[Fe/H] = -5.89 + 4.72  (V-I)
\end{equation}
	Using equation 1, the average metallicity of the clusters in our field is Fe/H=-0.74 dex, which is in close agreement with the value of Fe/H=-0.86$\pm$0.20 
derived by Lee \& Geisler (1993) and close to the value of Fe/H=-0.95 reported by Cohen, Blakeslee \&
Ryzhov (1998). The slightly higher metallicity of our sample is probably a 
sign of the color/metallicity gradient. The blue
clusters have a mean metallicity of Fe/H=-1.41 dex ( using V-I=0.95) while the
mean metallicity of the red clusters is Fe/H=-0.23 dex. 

\subsection{The Radial  Color Distribution And Its Implications On Formation Scenarios}
 
 The reason for the bimodality in the color distributions of the GCS of many elliptical galaxies is open to interpretation. A possible
scenario,  proposed by Ashman \& Zepf (1992) and Zepf \& Ashman (1993), is that  a merger  produces 
a metal rich population of globular clusters that is redder than the original 
metal poor population, thereby leading to a bimodality in the color 
distribution.  If the red clusters are created during a gas rich merger we might
 expect them to have formed closer to the center of the galaxy due to the higher degree of 
gaseous dissipation than suffered by the older blue population. This would 
 manifest itself as a negative color gradient with respect to galactic
radius. 
A recent paper by Forbes, Brodie \& Grillmair (1997) points out that the predicted correlation between increasing Sn and the fraction of red-to-blue
clusters predicted in this scenario (Zepf \& Ashman 1993)  does not hold for cD 
galaxies such as M87. With a Sn of roughly 16 we would expect 4 times more red
clusters than blue ones, while we find roughly equal numbers of red
and blue clusters in the inner region (The red clusters slightly outnumber the blue ones). This is an important result and probably indicates
that the simple equation spiral + spiral = elliptical is too
simplistic, especially for cD galaxies like M87 which have probably had
a more complicated formation history. Another possibility is that cD galaxies accrete dwarf galaxies 
and their cluster population (Zepf, Ashman \& Geisler 1995) or acquire
some of their globular clusters by   tidal stripping (Forbes, Brodie \&
 Grillmair 1997). This may partly explain the abundance of blue clusters 
since most of the accreted globular clusters will be metal poor and bluish in color. C\^{o}t\'{e}, Marzke, \& West (1998) on the other hand assume that the 
red clusters represent the galaxy's intrinsic population while the entire blue
 population is acquired through mergers of smaller galaxies or tidal stripping.
We discuss the relative merits of the various models in the following sections. 

	The issue of radial gradients in the color distribution of globular
clusters in M87 has been rather controversial. Though Strom {\it et al.} (1981) 
 reported that the average color of the clusters tends to be bluer at larger
galactocentric radii, later observations by Cohen (1988) and Couture {\it et al.} (1990)
 contested this claim. However, more recent observations of the region between 
50$''$$<$R$<$500$''$ by Lee \& Geisler
 (1993) and Cohen, Blakeslee \& Ryzhov (1998) have confirmed the radial trend
in metallicity (color) observed by Strom {\it et al.} (1981). 
 Interestingly,  Harris {\it et al.} (1998) have studied  the color distribution 
of clusters in the inner 
140$''$ of M87,  a field which is nearly identical to ours, and  conclude that color distribution is essentially flat within the core radius of $\sim$1$'$
  and then becomes bluer with radius in a manner that is consistent
with the gradients seen at larger radii. 

	In the  the top panel of Fig 11 we plot the  distribution of the 
globular cluster colors vs galactocentric radius for our data. The surface color
 distribution of the galaxy  within the inner 15$''$ as measured by us and the profiles from Zeilinger {\it et al.} (1993) are overlaid on the plot. The globular cluster color
 distribution  shows only a weak trend with galactocentric distance. A 
linear fit to the GC color vs distance for clusters with uncertainty of less
than 0.2 mag in V-I gives :
\begin{equation}  
V-I = 1.11(\pm0.01) - 3.5(\pm2.5)*10^{-4}*R
\end{equation}
where R is the
 distance in arcsecs from the galactic center. The small negative
gradient in the color distribution suggests that the mean metallicity (color)
of the clusters decreases with galactocentric radius. Even though the slope
derived in equation 2 is just a 1.4$\sigma$ result, the derived metallicity
 gradient is consistent with the Lee \& Geisler's (1993) data. In order to
 compare our observations with earlier results we calculated the mean color of the globular clusters in ten radial bins and the corresponding
 metallicity using equation 1. The metallicity of the clusters derived from this
dataset, Lee \& Geisler (1993) and Harris {\it et al.} (1998) are plotted as a function of logarithmic galactocentric radius in the bottom panel of Fig 11 (Note that we
used equation 1 to convert the Harris {\it et al.} colors to Fe/H and not the
 expression quoted by them  in order to be self consistent). Even though the 
derived metallicity of the clusters is highly sensitive to the photometric zeropoint and the coefficients of the conversion relation (equation 1), our 
metallicity estimates appear to agree well with previous observations.  Our 
data is also consistent with the trend of decreasing metallicity of  globular
 clusters with distance  observed by Lee \& Geisler (1993). Even though Harris
 {\it et al.} (1998) argue  that the mean 
metallicity within the core is constant, our observations (Fig 11) do 
not provide any 
compelling supporting evidence for this claim.  Given the 
uncertainties in the calibration of the metallicity scale, we opt to fit
a straight line  between log(R) and [Fe/H] in order to describe the metallicity gradient analytically. The
expression for the best fitting line is:
 \begin{equation}  
[Fe/H] = 0.06(\pm0.14)-0.44(\pm0.07)*log(R)
\end{equation}
where R is in arcsecs.

	 It is apparent that the mean metallicity of the clusters
decreases with galactocentric distance, but what is the underlying reason for this trend? One possibility is that the distribution of metallicities varies smoothly with galactocentric radius and the gradient simply reflects the metal enrichment of the infalling
gas associated with the collapse phase of the galaxy. The other explanation
 is that the bimodal metallicity distribution represents two distinct
cluster systems with different spatial distributions and that the difference 
in the relative number of the two sets of clusters causes the metallicity 
gradient. In order to search for evidence to corroborate either of these 
 hypotheses we have divided the  cluster population  into 5 radial bins, each 
having approximately one-fifth of the total clusters, and plotted the color distributions in Fig 12.  We see clear evidence of bimodality in each of the 
bins, with the blue and the red peaks located in same place in each of the 
figures.  A close look at the figures
 suggests that in the inner region there are more red clusters (V-I$>$1.09) than blue
 ones (V-I$<$1.09)	while in the outer region  blue clusters outnumber red ones. This trend is relatively weak and Kolmogorov-Smirnov tests show that it is statistically likely 
that all the distributions are identical except for the innermost one
at 20.9$''$ compared to the outermost sample at 82.6$''$.
 There is only a 0.33\%
 chance that these two distributions are identical.  More convincing evidence of this trend comes from WFPC2 observations of clusters in other fields around M87 (Elson \& Santiago
1996b, Neilsen {\it et al.} 1997, Neilsen {\it et al.} 1998). Neilsen {\it et al.} (1998)
have studied 4 fields in and around M87 including this dataset and the Elson
and Santiago (1996b) field, and find that in each case there is a blue peak 
near V-I=0.95 mag, while the red peak at V-I=1.2 mag becomes progressively 
weaker with distance. For comparison, we find approximately equal numbers of 
red and blue clusters in our field while Elson \& Santiago (1996b) found twice as many blue globular clusters as red ones in a field 2.5$'$ from the center of the galaxy. Based on the evidence on hand we conclude that there are two
 distinct populations of clusters in M87, with the red  clusters being
more centrally concentrated than the blue ones. Moreover, the radial trend in
 color is a natural consequence of the increasing ratio of blue to red clusters
with distance.   Recently Geisler,
Lee \& Kim (1996) have found that the red cluster population in NGC4472 is
similarly more centrally concentrated than their blue counterparts suggesting
that this phenomenon is fairly typical in giant elliptical galaxies.

	The Zepf \&
 Ashman (1993) merger model successfully explains most  of the features of the  color distribution described above.  However, we note that though the red
clusters slightly outnumber the blue ones in our field, the blue population
has a larger spatial extent, which suggests that overall there is a significant
number of both. Hence the overabundance of blue clusters also contributes significantly
to the high S$_N$ of M87.

 The
excess blue clusters  may have been acquired through cannibalization of metal poor satellite galaxies or by tidally stripping 
them of their clusters or possibily   the entire 
blue cluster distribution is a cannibalized population as suggested by C\^{o}t\'{e} {\it et al.} (1998). Another possibility is the Harris {\it et al.} (1998)
 scenario according to which the
blue clusters were formed during the collapse phase of a massive proto-galaxy
and supernova  powered galactic winds drove out a large portion of the
gas, leaving behind a blue cluster rich  galaxy. The extended spatial distribution and overabundance of 
blue clusters predicted by this theory would also agree with our observations.

  We also observe that the mean population is  0.2 mag bluer than the stellar
background. This is much smaller than the difference of 0.5 mag between the
mean cluster color and the galaxy background reported by Couture et 
al. (1990) using the Boroson {\it et al.} (1983) value of V-I = 1.6 mag for the stellar background. We believe that this large difference may partly be due to
to the fact that the Couture {\it et al.} (1990) I band photometry is based on the Kron-Cousins
system while Boroson {\it et al.} (1983) reported their I magnitudes in the Gunn system (see $\S$2.5).
 The observed difference of only 0.1 mag between the red clusters and the galaxy background weakens the argument of Couture {\it et al.} (1990) that the GCS formed much earlier epoch than the
 bulk of the stars and strengthens the case for a second burst of cluster
 formation late in the metal enrichment history of the galaxy, possibly
due to a merger.

\subsection{ The Effect Of Color On The Luminosity Function }

   Ashman, Conti \& Zepf  (1995) (henceforth
ACZ) have modelled the luminosity of globular clusters with different metal 
abundances and they find that the absolute magnitude of the peak of the GCLF
should vary with metallicity if we assume identical mass functions.  Therefore, according to ACZ, bluer, metal poor clusters  should be 
brighter than  redder, metal rich clusters in the V band due to the effects of metallicity on stellar evolution. 
   In Paper 1 we found that the blue population is 0.13 mag   brighter  than
 the red ones in the V band, consistent with the sense of the prediction although 
smaller in magnitude. Elson \& Santiago (1996b) 
also reported a difference of 0.3 magnitude  in the same sense between the two 
populations of clusters in their sample.  We find that the blue clusters 
are 0.23 mag brighter than the red ones in the V band for the candidate clusters
 used in this paper. The blue
 population  identified by a homoscedastic partition in 
$\S$ 3.3 is 0.30 mag brighter than the corresponding red population in V, while the
difference is 0.22 mag for the heteroscedastically partitioned set.
 However, the magnitude of the difference is still smaller than
the value of $\approx$0.5 mag  predicted by ACZ. In
 the I band, ACZ predict that the blue population should be brighter than the
 red population by  $\approx$0.1 mag. However, the I bands magnitudes show
a very small trend in the opposite sense i.e. we find that the red clusters
are brighter than the blue ones by $\approx$0.06 mag. This apparent
 inconsistency may be a result of the simplifying assumption made by ACZ that
both the populations are the same age. By relaxing this criterion, and allowing
the  age of the metal rich population to be younger, the colors and magnitude
shifts in both bandpasses can be explained self-consistently. Using the
 evolutionary tracks of Whitmore {\it et al.} (1997) derived from the Bruzual \& Charlot (1998) models, we estimate that the red clusters are 3-6 Gyr younger than a 
15 Gyr old blue population. A similar analysis based on the Worthey (1994) models gives an age of
9 Gyr for the red clusters. The 9-12 Gyr age of the red clusters supports the 
hypothesis that they were formed during a second event, later in the history
of the galaxy than the blue clusters.

	We have established previously that the luminosity function of the
entire cluster distribution shows no clear evidence of radial variation, but the
red and blue cluster distributions separately might show a spatial variation.
In order to  address this question we, once again,  divided the clusters into five radial bins with approximately
equal number of objects and then divided them into red and blue populations
using the mean value of 1.09  as the breaking point. We calculated the luminosity function in each radial bin, and the difference in the luminosity 
functions of the red and blue clusters. We found that the slope of the straight line that best fits the difference in the luminosity function of the red and blue clusters with magnitude, in each case, is either statistically zero or 
a very small  negative number (i.e. $\approx$1 $\sigma$ result) suggesting
that the blue clusters are slightly brighter than the red ones. On the whole
 the red and blue clusters seem to have very  similar luminosity distributions everywhere. In order to illustrate this we present  the mean magnitudes 
for clusters  brighter than 23.5 mag in V, where incompleteness is not
a factor,  in Table 3

  The blue clusters also shows a slightly smaller statistical variation in 
magnitude with
 radius than the red clusters suggesting that they 
 might be a more stable  indicator of the 
turnover luminosity than the combined population. Such small scale variations 
notwithstanding, the most significant result is that on the whole the 
luminosity distribution is remarkably constant everywhere.

\subsection{ Surface Density of Clusters}

	The core radius of the  globular cluster system of M87 has been   
shown to be larger than that of the underlying galaxy in earlier 
studies by Grillmair {\it et al.} (1986), 
Harris {\it et al.} (1986) and Lauer \& Kormendy (1986). In Fig 13 we plot the logarithmic
 surface density of clusters for the various datasets  as a function of logarithmic radius. Comparison of the underlying galaxy's brightness 
profile with that of the cluster density distribution confirms previous
observations that it's profile  is much  flatter than that of the galaxy 
light in the
 inner region of the galaxy.  
 The HST's superior ability 
to identify globular clusters near the center of an elliptical galaxy is also immediately
 apparent from the figure. 	

	The density profiles plotted in Fig 13 are offset from one another since the 
datasets have varying completeness limits. In order to calculate the corrected density profile of the cluster system, we calculated  the projected total number of clusters at each point  assuming that the luminosity function at each point 
has the same turnover and halfwidth as the entire population (Note that we calculate the total number of clusters that should be visible if clusters everywhere follow the GCLF plotted in Fig
 5 and if the completeness was 100\% everywhere). A similar correction was made to the Grillmair {\it et al.} (1986)
dataset after applying a color correction of B-V=0.67 mag (Couture {\it et al.} 1990).
We did not use the other datasets to calculate the core radius because 
the completeness limit is not well defined for the Harris (1986) dataset 
and  is much lower for the Lauer \& Kormendy (1986) data.
 The projected total density distribution  is  plotted in Fig 14. The best fitting King model (King 1966) with a concentration
 index of  2.5 is  overlaid  for comparison. The King radius, r$_0$ (sometimes referred to as the core radius), derived from the fit   is 56.2$''$ and  is much larger than  the reported core
radius of the galactic light (R$_c$=6.8$''$ Kormendy 1985). Though our 
GCS core radius  is  smaller than the  88$''$ reported by Lauer \& Kormendy
(1986), it is consistent with the $\approx$1$'$ value derived by McLaughlin (1995). 

 Though destruction 
of clusters due  to tidal shocking and  dynamical friction could have conceivably caused this turnover in the density distribution, the fact that we see no clear spatial change
 in the luminosity function ($\S$ 3.2) makes it  unlikely. We agree with previous studies
(Grillmair {\it et al.} 1986; Harris {\it et al.} 1998 etc.) that conclude that the spatial 
constancy of the GCLF suggests that the large core is a
 relic of the cluster formation process with  the initial  distribution of clusters being less peaked than the underlying galaxy's light profile. If the 
blue clusters were formed during the collapse of a huge proto-galaxy as 
suggested by Harris {\it et al.} (1998) it is possible that they mimic the density 
distribution of the mammoth proto-galaxy and hence the discrepancy in the
core radii. Similarly, if the population of blue clusters is entirely 
cannibalized, as suggested by C\^{o}t\'{e} {\it et al.} (1998) the cluster distribution would be predicted to have a large core radius. In a recent HST study of 
14 ellipticals with kinematically distinct cores, Forbes {\it et al.} (1996) 
found that the  globular cluster density distributions rose less steeply than
the galaxy background in the nuclear region of all 14  galaxies. They discounted
destruction processes being the cause of this turnover for the same reasons 
(i.e. the luminosity function is similar everywhere).

	The projected central density of clusters is 460 clusters arcmin$^{-2}$.
For comparison Lauer \& Kormendy calculated a central density of 72 clusters 
arcmin$^{-2}$ from their observations. This large difference in densities is 
another reminder of  the HST's superior ability to resolve point sources 
in a strong background.

 \subsection{Shape of the cluster distribution}
   
	The shape of the globular clusters distribution may hold some important
 clues about the formation  history of the galaxy since the spatial distribution of the clusters retains information about the shape of the proto-galaxy from which it
formed  and/or the signatures of violent interactions that may have changed the morphology of the galaxy. McLaughlin {\it et al.} (1994) studied the spatial distribution of the 
globular clusters around M87 in great detail and concluded that the cluster 
distribution is elliptical ($\sim$E2.5 Hubble distribution) and aligned with the major axis of the galaxy 
in the region 1.9$'$$<$R$<$4.5$'$.  

In the top panel of Fig 15 we plot the number of cluster candidates within 55 arcsecs 
of the center of the galaxy, binned in 30$\arcdeg$ sectors, as a function of angle East from North. Note that we have used reflection around the 
center of the galaxy to complete the L-shaped wedge that is not covered by the
PC.  Although the isophotes of the galaxy in this region are nearly circular,
we can clearly see that the spatial distribution of 
globular clusters is flattened. Histograms of the red and blue cluster distributions show that they are both flattened with the red clusters 
having a larger ellipticity than the blue ones. We have also plotted the actual positions of the red and blue clusters on the WFPC2 chip in the lower panels of
Fig 15. 

 In order to quantify the ellipticity of the 
clusters distribution we created a fake image in which we added a 
Gaussian source at the location of each cluster brighter than V=23.5 (in order
to minimize the effects of incompleteness), and   then smoothed it by 
convolving with a wide Gaussian of FWHM $\approx$ 30$''$. At a distance of 
55$''$ from the center of the galaxy (major axis), we find that while 
 the blue clusters follow  an E2.8$\pm$0.2 (Hubble type)
  profile, the red cluster distribution is flatter and follows an 
E3.6$\pm$0.2 shape. One concern is that
smoothing of the fake image by a broad Gaussian may lead us to significantly 
underestimate the ellipticity of the clusters distribution.  Numerical
tests described in Kundu \& Whitmore (1998) indicate that this is a minor effect in the case of M87, hence we estimate that the shape of the 
blue cluster distribution is E3$\pm$0.5, while that of the red clusters is 
 E4$\pm$0.5. We also find  the position angle of the major axis of the red clusters is 185$\pm$5 East of North and the position angle of the blue clusters
is 195$\pm$5 East of North. 

	The fact that both the red and the blue clusters have flattened distributions that are roughly coincident, while the galaxy light profile is circular, is an intriguing result since  no formation or destruction mechanism associated with the present day
spherical halo can induce the  observed shape of the  clusters. The elliptical 
profiles of the clusters must then be a signature of the shape of M87 at an earlier epoch  when it was  flatter, maybe even with a disk
component. If the cluster distribution of M87 is largely formed by accretion of 
companions, the elliptical shape may reflect the shape of the Virgo cluster 
itself. The discovery that the position angle of cD galaxies
correlates with that of the host galaxy by Binggeli (1982) seems to support this hypothesis. 

	We also note that the position angle of the the major axes of both the blue and red clusters differ from the position angle of the major axis of the galaxy,
 which is at 155$\arcdeg$, by a statistically significant amount. This is in direct contrast to the observation of McLaughlin {\it et al.} (1994), who reported 
that the major axis of the clusters in the region 1.9$'$$<$R$<$4.5$'$ is aligned
with the major axis of the galaxy. The cluster distribution apparently 
has twisted iso-density curves. Intriguingly the position angle of the clusters 
within our field of view, especially the red ones, seems to coincide with the
181$\arcdeg$ position angle of the nuclear ionized disk. While it is 
speculative 
to link the large scale structure of the globular cluster with a 1$''$ radius 
ionized disk,  it is possible that the nuclear disk is actually a remnant of an
 accretion event that produced many of the central red clusters.    

\subsection{ The Size Distribution}

 As we noted in the aperture photometry section, the profiles of the globular clusters are on average broader than stellar 
profiles, indicating that they are spatially resolved.
	In order to model the light distribution of the globular clusters
in M87 we assume that they are similar to the Milky Way 
 clusters and that the surface brightness profile can described by 
 King models (King 1962). The size of a cluster can then be defined  by two 
parameters, the King radius (r$_0$), and the 
concentration parameter c = log$_{10}$($\frac{r_{tidal}}{r_0}$). However,
there are two important considerations to be made before we model  the observed light
distribution of the clusters. We must take into account the fact that the
WFPC2 point spread function varies across each of the chips. Also,
 the undersampling of the PSF by the WFPC2 camera has 
the unpleasant effect of modifying the observed light profile
depending on the location of the peak  within a pixel.

	We created 4-fold oversampled Tiny Tim models (Krist 1995) of the PC and
 WF PSFs at five  evenly spaced locations on each of the chips.
Each of these PSFs were convolved  by a range of King models, varying c between 0.5 and 2.5, 
and r$_0$ between 0.5 and 16 pixels of the oversampled PSF. We then resampled the models to normal size 
for eight different locations of the peak within a pixel. For each of the
models, we performed photometry for a range of aperture radii to create 
a cumulative light profile and normalized all the profiles in our library to 
the observed light within 3.3 pixels. The profiles of  the candidate objects 
were created using the same aperture photometry parameters.  We  deduced the
structural parameters of an individual cluster by  finding the
best fit (in a  least square sense) within the set of models that were closest
to the  candidate cluster with respect to the chip location and centering 
within a pixel. As explained in Kundu \& Whitmore (1998), our numerical tests 
indicate that there  
are significant correlated errors  when we fit c and r$_0$ simultaneously. On the other hand, if we restrict the fits to a single 
concentration index, we can measure the relative sizes of the 
clusters candidates reliably even in
 regions with a strong galaxy background. We quantify the sizes in terms of the  physically meaningful half light radii (r$_h$) because unlike the King radius,
r$_0$, it is largely unaffected by the choice of the King model c parameter. We chose to fit c=1.25 and c=1.5 King models since the median King model concentration parameter of old globular clusters in the Milky Way (Harris 1996), M31 (Crampton {\it et al.} 1985; Grillmair {\it et al.} 1996), and the LMC 
(Elson \& Freeman 1985) all lie within this range.  

The half light distribution of the clusters, r$_h$, obtained by fitting the cluster profiles to c=1.25 King model convolved PSF,  is plotted as a function of the cluster brightness 
in Fig 16. While we  see no obvious relationship between r$_h$ and V, we note 
the striking resemblance of our plot with Fig 1 of van den Bergh (1996) which
plots the same quantities for the Milky way globular clusters. As in the 
Milky way, the clusters with half light radii between 2 and 4 pc are more luminous than both smaller and larger clusters. The lack of any strong correlation between the sizes and the brightness (mass) of the clusters also
implies that larger clusters are in general more diffuse, low density objects.

	Fig 17 is a plot of the size  distribution of the clusters as a 
function of the projected distance from the center of M87. We see no significant radial trends in the figure. However, a careful inspection of the plot shows that there may be a lack of large 
clusters within the innermost 10$''$ and  the 
large, diffuse clusters in the innermost regions are destroyed by tidal forces.
Given the small number statistics we cannot conclusively prove whether  
this is indeed a real effect.

The  mean half light radii (r$_h$)
of the cluster candidates in the PC and WF, fitted to  c=1.25 and c=1.5 models, are shown in
Table 4.  The measured value of $r_h$$\approx$2.5 pc is comparable to the mean 
half light radius  of $\approx$3 pc for the brightest Milky Way clusters (van den 
Bergh 1996).  Table 4 also shows the average sizes of the blue and red clusters in the PC and WF chips. Interestingly, we find that the blue clusters are statistically 
larger than the  red clusters in both the WF and the PC.  
  We have observed a similar effect in the NGC 3115 globular
cluster system (Kundu \& Whitmore 1998). In order to verify the reliability of the size difference we developed a different
algorithm  to  estimate the size of the clusters.  First, we  obtained  counts in a set of radial bins around the candidate objects and calculated the FWHM of  Gaussian curve that best fit the  intensity distribution 
of each individual cluster.  WFPC2 observations of a  set of  program
stars from a  field in $\Omega$ Cen were then convolved with Gaussian 
distributions of varying widths and fitted with the same routine to 
establish a relationship between the convolved width and resulting FWHM.
 The calculated sizes of the blue clusters in M87 are larger
than the red clusters in all the chips, in both the V and I filters using
this method.

 We tested 5 independent star fields in $\Omega$ Cen to see whether this  might be an instrumental effect. In each of the test images we selected unresolved/barely resolved objects with colors between 0.8$<$V-I$<$1.4 and used our algorithms to determine the size. We find that the size of objects with V-I$>$1.06 mag is statistically 
identical to the size of objects with V-I$<$1.06 in both the V and I filters. We also tested each
 quadrant of all four WFPC2 chips and discovered no position dependent effect 
that may be inducing the difference in size between red and blue objects.  We therefore conclude that the observed size difference between the
red and blue clusters  is  real and not due to instrumental biases. 

 van den Bergh (1994) showed  that the half-width of the Milky Way globular clusters
increases with galactocentric distance. Since the blue clusters are on average more distant from the center of the
galaxy than the red disk clusters, the difference in size between the red and
the blue clusters may be indicative of a similar relationship in NGC 3115 and
M87.  The physical reason for the difference in sizes between the red and blue 
clusters remains unclear. It may  be  a relic of the different formation mechanisms 
of blue and red clusters, a signature of the radial variation in efficiency of various destruction mechanisms superimposed on the different spatial scales
of the two systems, a reflection of slightly accelerated core collapse of the 
red clusters which are closer to the center of the galaxy, or a combination of 
all of the above.

\section{Summary}
 
	We observed  the inner region of M87  with the WFPC2 camera onboard the 
HST and identified 1057 globular cluster candidates. These observations
 reached two magnitudes deeper than the turnover magnitude of the
 GCLF and allowed us to study the luminosity, color, and size distribution of the globular
clusters in the inner region. The main results gleaned from this study are:

1)  Within statistical errors, we find no variation in the luminosity 
function with radius, in either the V, or the I band, hence no
 obvious evidence of 
evolution of the  luminosity function due to destruction processes. The
 constancy of the GCLF bolsters our confidence  in the turnover magnitude 
 as a secondary 
distance indicator.  Additionally,  I band magnitudes show slightly less scatter than the
 V band values presumably since the magnitudes 
are less affected by  metallicity differences. This suggests that there might be
 some merit in using I band GCLFs instead of V band GCLFs as distance indicators.

2) The peak of the turnover magnitude  is  m$_V$$^0$=23.67 mag in the V band and
m$_I$$^0$=22.55 mag in the I band.
 
3)	The color distribution of the clusters is bimodal with a
blue peak at V-I = 0.95 mag, a red peak at V-I = 1.20 mag and a mean color 
 V-I = 1.09 mag. The bimodality in the color distribution  reflects the underlying  bimodal metallicity distribution of clusters in M87. The blue
peak has a metallicity of Fe/H=-1.41 dex while the more metal rich, red peak
has a metallicity of Fe/H=-0.23 dex.   The difference in the color of the red population
 and the underlying galaxy is only 0.1 magnitude, smaller than the 0.5 magnitudes
previously reported in the literature, which suggests that the red population 
may have formed at 
a fairly late stage in the metal enrichment history of the galaxy, probably during a metal rich merger event. 

4) The color distribution is bimodal 
in all five radial bins and there is weak evidence that the red clusters
are more centrally concentrated than the blue ones within our WFPC2 field of
view. Combining our data with other observations,  we infer that the average
 metallicity of globular clusters decreases with distance, most likely due to 
the increasing ratio of blue to red clusters with galactocentric distance. 

5)  The luminosity function for  the red and blue
 clusters are  similar at all radii, though the blue candidates are on average 0.2 magnitudes brighter than the red ones in the V band. The difference in the color and brightness between the two populations suggests that the red clusters were formed roughly 9-12 Gyr ago, assuming a 15 Gyr old blue population.

6)	The core radius of the globular cluster density distribution is  56.2$''$ for the best fitting King model with a concentration index c=2.5. This  is much larger than that of the underlying galaxy light (R$_c$=6.8$''$).  Since we 
see no evidence of cluster destruction processes in the luminosity function,
this is most likely  a relic of the formation history of the clusters.

7) Even though M87 has a $\sim$E0 shape, the globular cluster distributions
appear to be flattened. While the blue cluster population has an E3$\pm$1
 profile the red clusters have an even flatter E4$\pm$1 shape. This implies that
 M87 may have had a much flatter profile during the epoch in which the clusters formed.

8)	The half light radius of the clusters is $\approx$2.5 pc with
no obvious radial variation in the size distribution. On average, the 
blue clusters appear to be 20\% larger than the red clusters.

	We conclude from this study that the globular clusters in the inner
 region of M87 have remarkably homogeneous spatial properties and that the Globular
Cluster Luminosity Function and  color distribution of the clusters 
are similar throughout the regions studied by us. However, small differences
in the properties of the red and blue clusters suggest that these two 
populations might have different formation histories and that the red population
  formed  later in the metal enrichment history of the galaxy than the 
blue population, most likely during a major merger. 

We are grateful to Alberto Conti for helping us with the KMM mixture modelling.
We would also like to thank Ivan King for supplying us with King model
 profiles and Bryan Miller for help in the early stages of the project. We also
 wish to thank the anonymous referee for many useful comments and suggestions.

\newpage
\begin{figure}
\caption{Mosaic of 4x600 sec V (F555W) band WFPC2 images of the inner region of M87.  Most of the
 point sources visible in the image are globular clusters. The arrow points 
North, the bar points East, and each is 5$''$ in length. \label{fig1}}
\label{key}
\end{figure}

\begin{figure}
%\plotfiddle{image3.ps}{7in}{0}{90}{90}{-300}{-120}
\caption{Median subtracted image of a portion of Fig 1. The optical jet
 and numerous globular cluster candidates can be seen right to the core
of the galaxy. The 100 cluster
candidates closest to the galactic center, identified in Table 2, are
 marked by circles. Arrow pointing North is 2.5$''$ in length \label{fig2}}
\end{figure}
\newpage

\begin{figure}
\caption{ Completeness curves for globular clusters in the V (F555W) image as a 
function of radial distance (See text for details).   \label{fig3}}
\end{figure}
\newpage

\begin{figure}
\caption{ Luminosity function for objects detected in a test field 
9.3 degrees from the source. The sources have been detected using the 
same automated routine that was used to detect candidates and remove contaminating
objects from the program field.   \label{fig4}}
\end{figure}
\newpage

\begin{figure}
\caption{ The Globular Cluster Luminosity Function for the full sample
in the V image. The bottom panel is a plot of the Luminosity Function
derived from the objects identified in Paper 1 using aperture corrections
for point sources and background and completeness limits calculated in 
Paper 1. The top panel is a plot of the GCLF derived from the objectively 
identified set using the aperture corrections in table 1 and completeness and 
background corrections plotted in Figs 4 and 5. The dotted lines give the 
background and completeness corrected distribution and the dashed lines give the 
best Gaussian fit. The turnover magnitude  for the objective set
is slightly brighter primarily because of the new aperture corrections   \label{fig5}}
\end{figure}
\newpage

\begin{figure}
\caption{ The radial distribution of the V
magnitudes of candidate clusters with photometric errors less than 
0.3 mag. Clusters near the center of the galaxy appear to be brighter than 
the more distant ones due to the fact that the completeness limit is at brighter
magnitudes closer to the center of the galaxy. \label{fig6}}
\end{figure}
\newpage

\begin{figure}
\caption{  The  sample has been
divided into 5 approximately equal radial bins and the luminosity function for 
each of the bins in both the V and the I is plotted. The dashed line shows the 
completeness corrected and background subtracted distribution. The GCLF has 
a similar shape in all the bins with no obvious differences that can be attributed to cluster destruction mechanisms. 
\label{fig7}}
\end{figure}
\newpage

\begin{figure}
\caption{ Parameters of the Gaussian curves fitted to the luminosity functions
in Fig 7. The bottom 
panel plots the turnover luminosity in V and I as a function of distance.
The top panel is a plot of the corresponding values of $\sigma$, the dispersion,
 for the Gaussian fits.  Both the turnover magnitude and $\sigma$ are constant within uncertainties indicating that the GCLF does not vary with distance. 
   \label{fig8}}
\end{figure}
\newpage

\begin{figure}
\caption{The cumulative profile of the corrected luminosity distributions in 5 radial 
bins normalized to 150 counts:  Only clusters brighter than 24 mag in V, and
 23 mag in I, have been considered  in order to limit the uncertainties due to  completeness and background correction.
Since the normalized  profiles in neither the V, nor the I band show a
 consistent, distance dependant variation, we conclude that there is no
 clear evidence of the effect of destruction mechanisms in the luminosity
function of M87. \label{fig9}}
\end{figure}
\newpage

\begin{figure}
\caption{ Histogram of V-I values of the clusters with measurement uncertainties less than 0.2 magnitudes. The bottom panel shows the histogram for the objects 
identified in Paper 1, whose brightnesses were derived  using point source 
aperture corrections from Paper 1. The top panel 
shows the histogram for the objectively identified set whose brightnesses were 
calculated using the new aperture corrections shown in table 1. The bimodal 
distribution of colors can be seen in both sets. Even though the V and I 
magnitudes  derived using the new aperture corrections are smaller, the color
distribution remains nearly identical. \label{fig10}}
\end{figure}
\newpage

\begin{figure}
\caption{ The upper panel plots the variation of the color distribution with radial distance. All cluster
candidates with a photometric error less than 0.2 mag in V-I are plotted
as a function of distance. The filled circles between 0-15$''$ plot the surface 
V-I magnitude per square arcsec of the galaxy light derived from our 
data. The dashed line gives the V-I galaxy profile measured by Zeilinger {\it et al.}
(1993). The lower panel is a plot of the variation in the  mean metallicity of the clusters as a function of radius. The average color of the clusters decreases with R. The dashed line represents equation 3, which describes the best fitting straight line  through the data. \label{fig11}}
\end{figure}
\newpage

\begin{figure}
\caption{  The 
 V-I histograms  at  5 radial distances. We see evidence of bimodality at all
 radii.\label{fig12}}
\end{figure}
\newpage

\begin{figure}
\caption{The surface density of the globular cluster distribution of
 M87 derived from the raw number counts. The values
obtained in previous studies are also plotted for comparison. The superior 
ability of the HST to detect clusters in regions of strong background is
 immediately apparent. The dashed line is the galaxy brightness profile from 
de Vaucouleurs and Nieto (1978) shifted arbitrarily in the Y direction by 
setting $\mu$$_B$=2.5 equal to Log $\sigma$ = 1.0, where $\sigma$ is the
number of clusters per square arcmin. The solid line is
the V profile from Zeilinger {\it et al.} (1993). It has been shifted by the same 
arbitrary amount assuming that B-V = 1.0 mag for the galactic light of 
M87. \label{fig13}}
\end{figure}

\begin{figure}
\caption{ The  projected total surface density of the globular cluster distribution in the inner region of 
 M87. The density has been calculated by assuming that the GCLF at each point is identical to the overall luminosity function. The solid line is the best 
fitting King model with concentration index of 2.5. The dashed line is the galaxy brightness profile from 
de Vaucouleurs and Nieto (1978). It is apparent from the figure that the
core radius of the  GC density profile is larger than that of of the underlying 
halo light distribution.\label{fig14}}
\end{figure}
\newpage

\begin{figure}
\caption{ The top panel plots the histograms of the cluster position angles within 55 arcsecs of the center of the galaxy. 
The two bottom panels show the locations of the red and blue clusters within the
 WFPC2 chip. Both the blue and red cluster populations are  flattened; The spatial distribution of red clusters appears  to be  flatter than that of the blue ones. The 
major axis of the cluster distributions are roughly aligned with the nuclear
disk which in turn is approximately perpendicular to the jet (marked with a 
bar in the lower panels)     \label{fig15}}
\end{figure}
\newpage

\begin{figure}
\caption{ The half light radius vs brightness distribution of the  cluster 
candidates. There is no clear relationship between the size and mass (brightness) of the clusters. As in the Milky Way, clusters with half light radii between 2 and 4 pc
appear to be brighter than smaller or larger clusters.  \label{fig16}}
\end{figure}
\newpage

\begin{figure}
\caption{ 
 The radial distribution of the sizes of the clusters as a function of distance.
 No obvious trend in the sizes is seen except perhaps a weak indication of a lack of large clusters
in the inner 10$''$. This may be caused by the tidal disruption of diffuse
clusters in the region close to the nucleus of the galaxy. \label{fig17}}
\end{figure}
\newpage

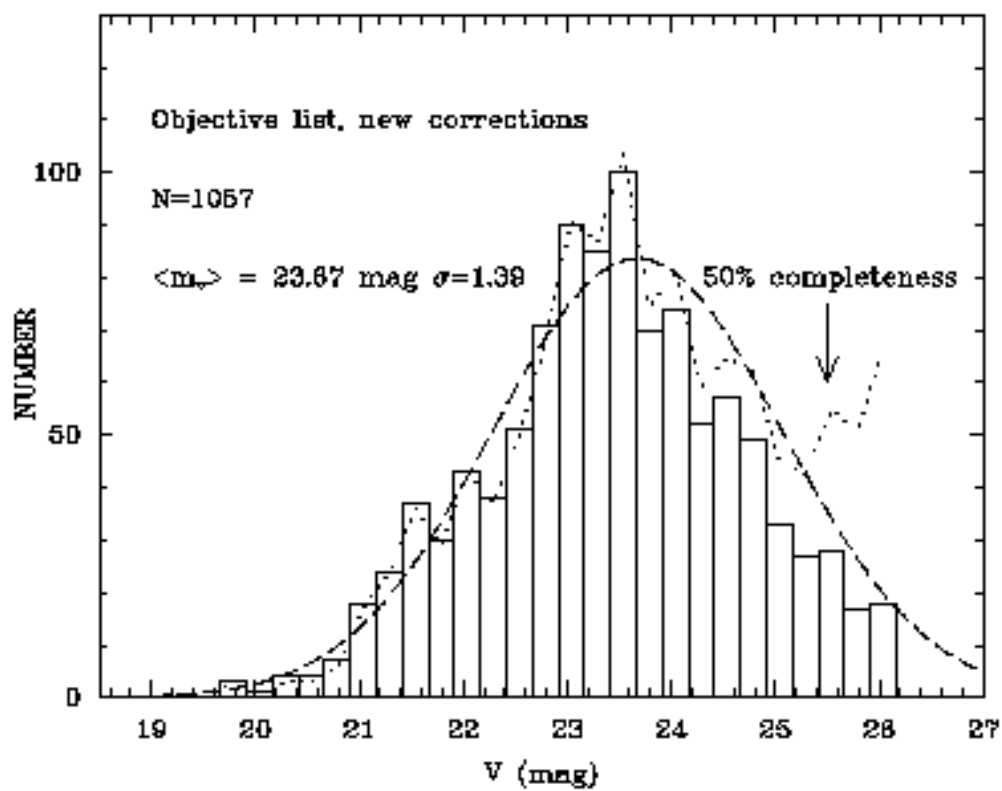
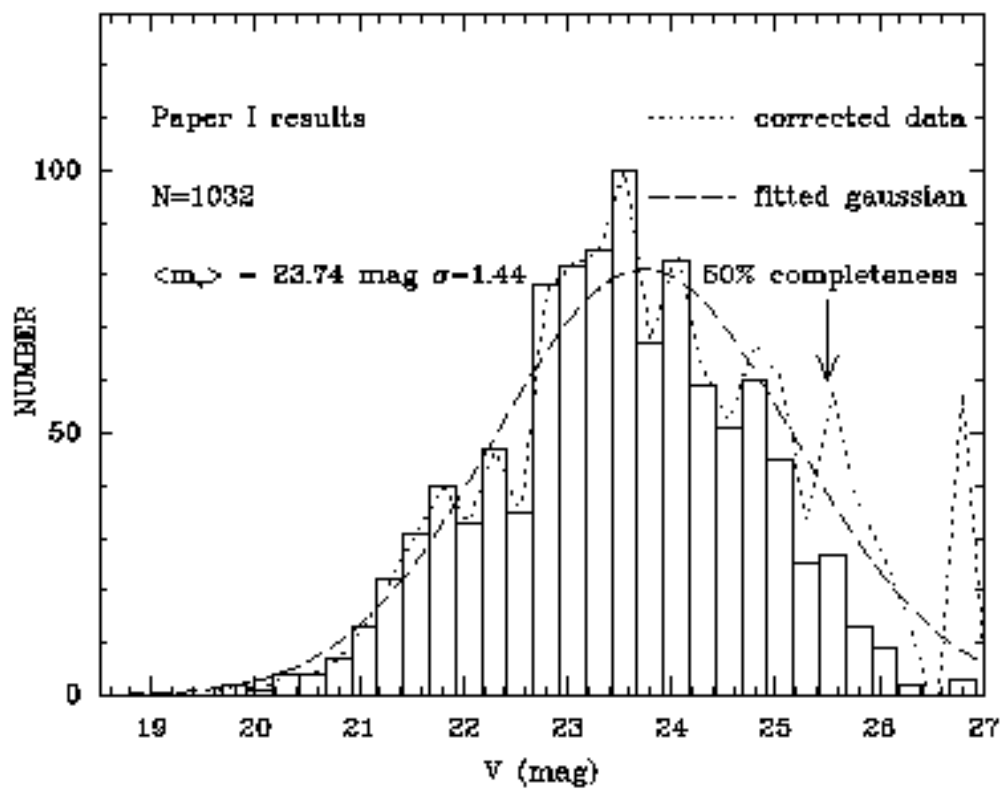
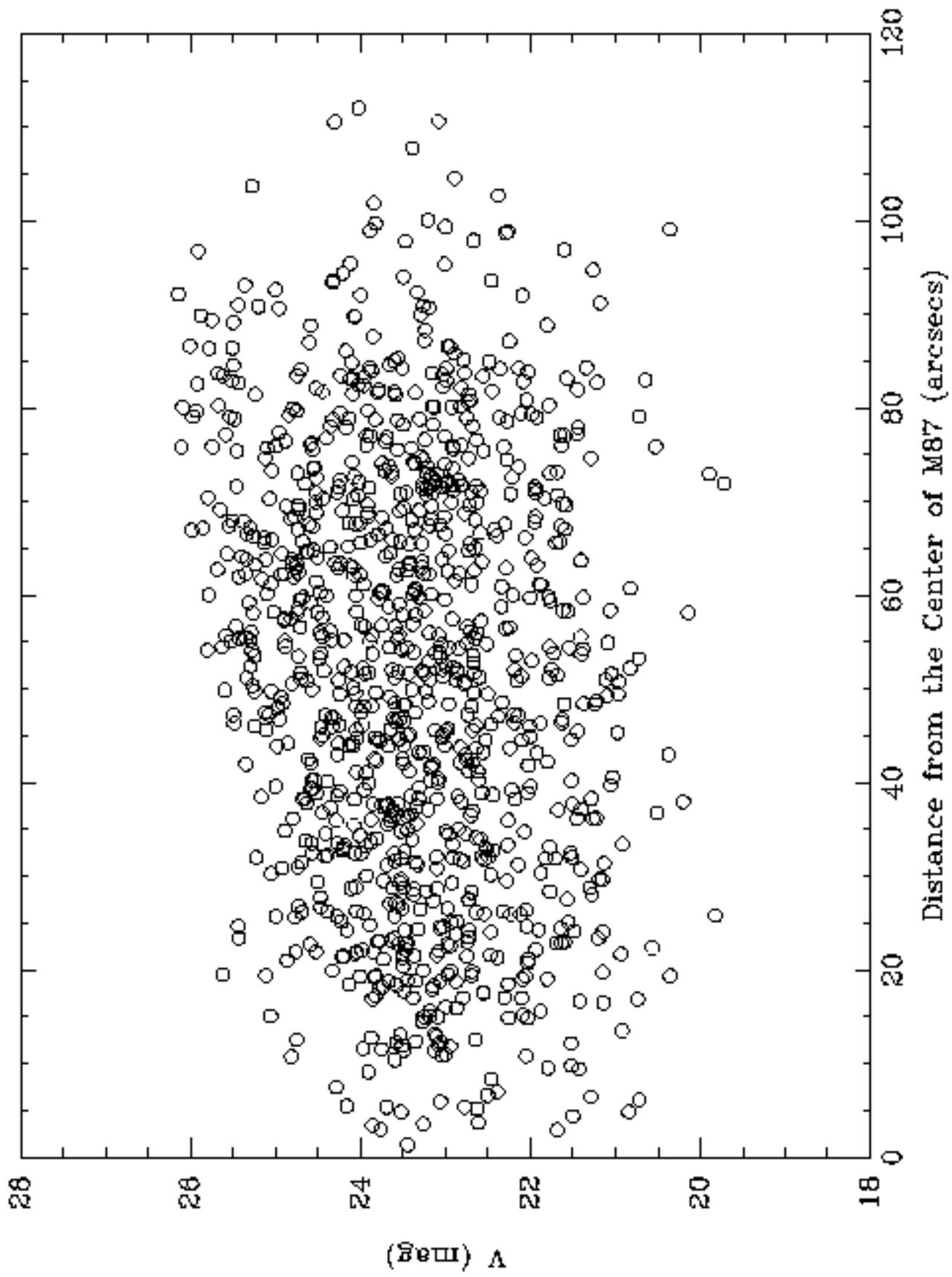
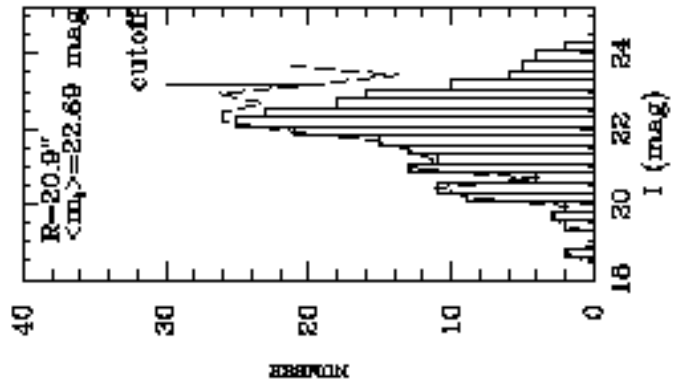
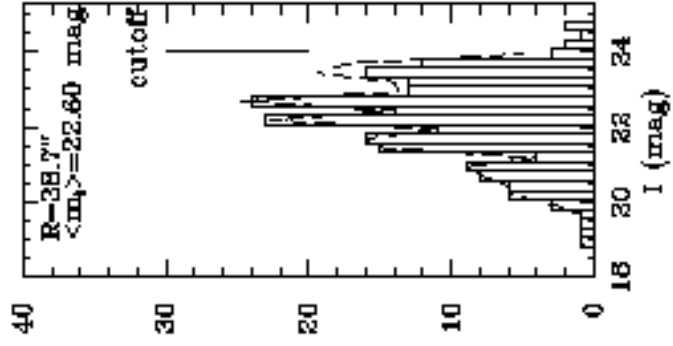
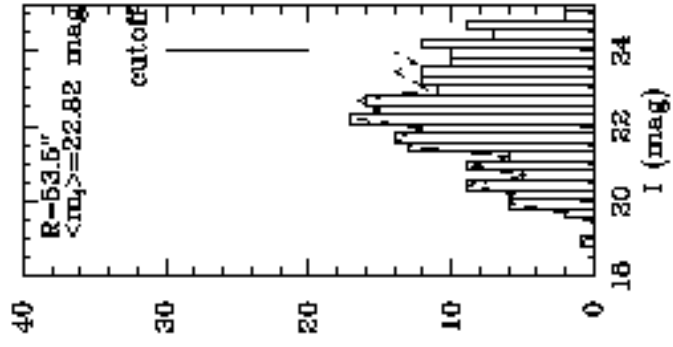
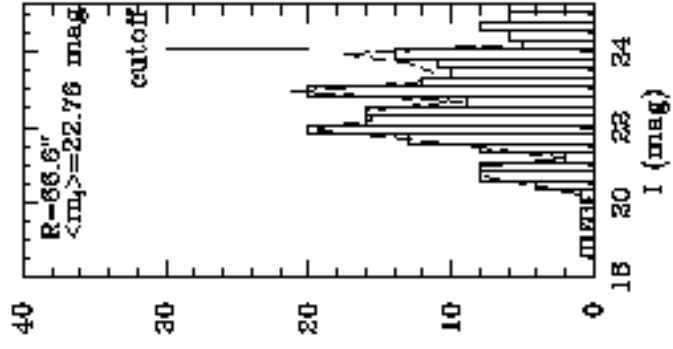
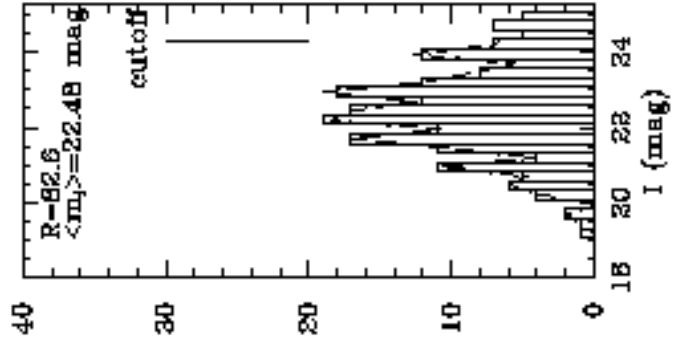
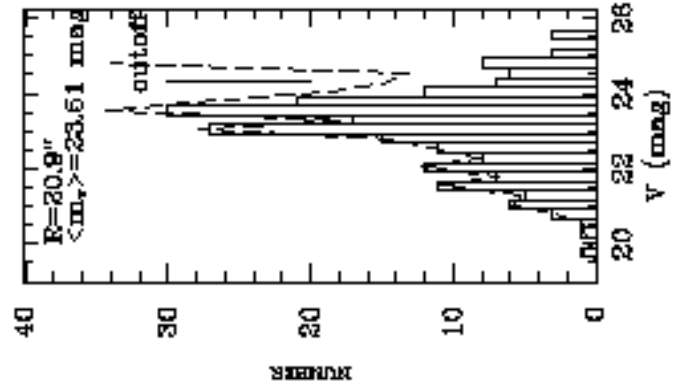
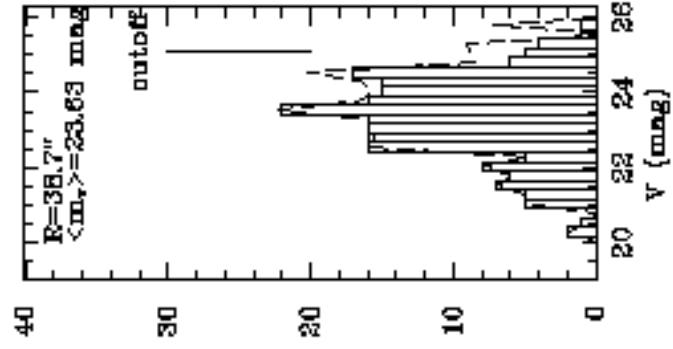
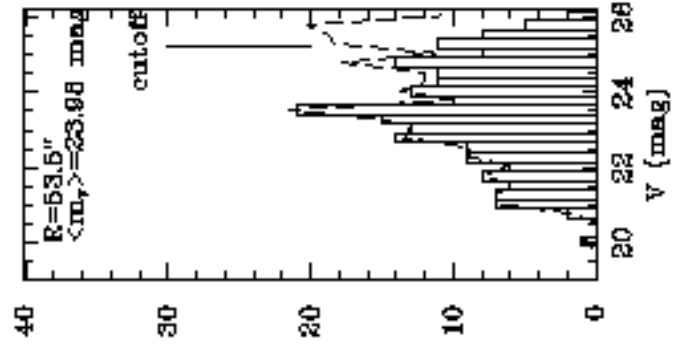
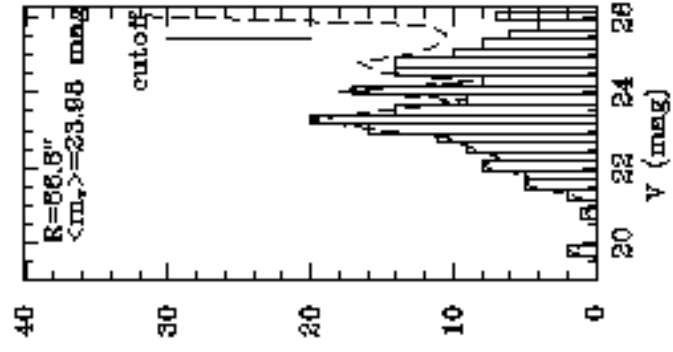
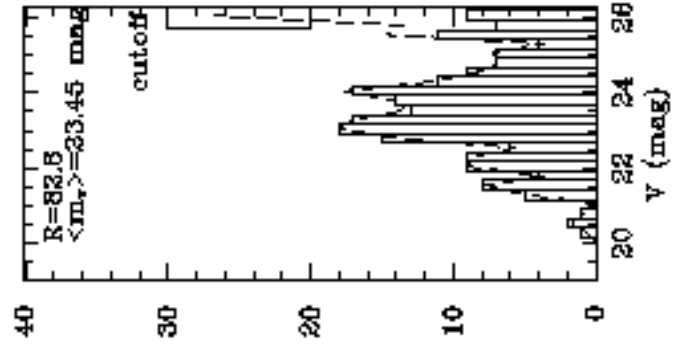
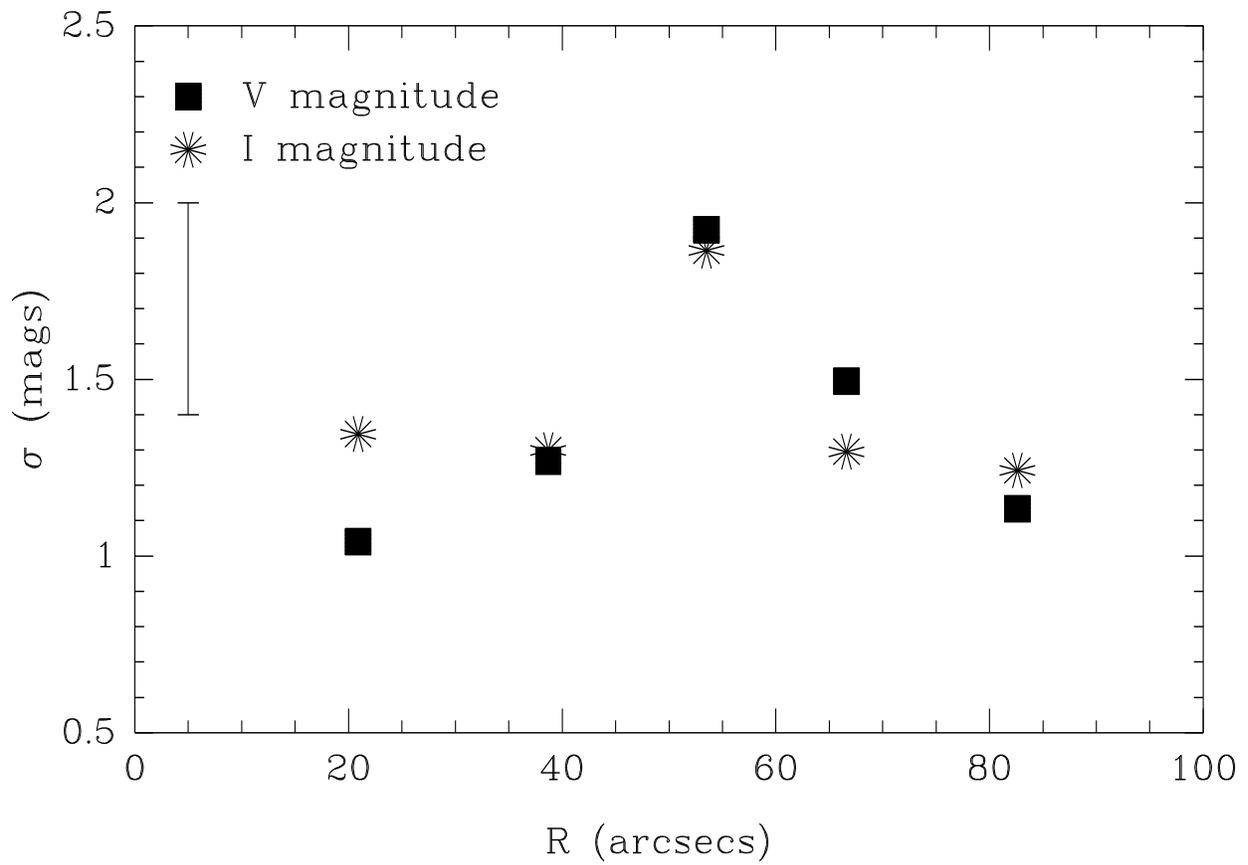
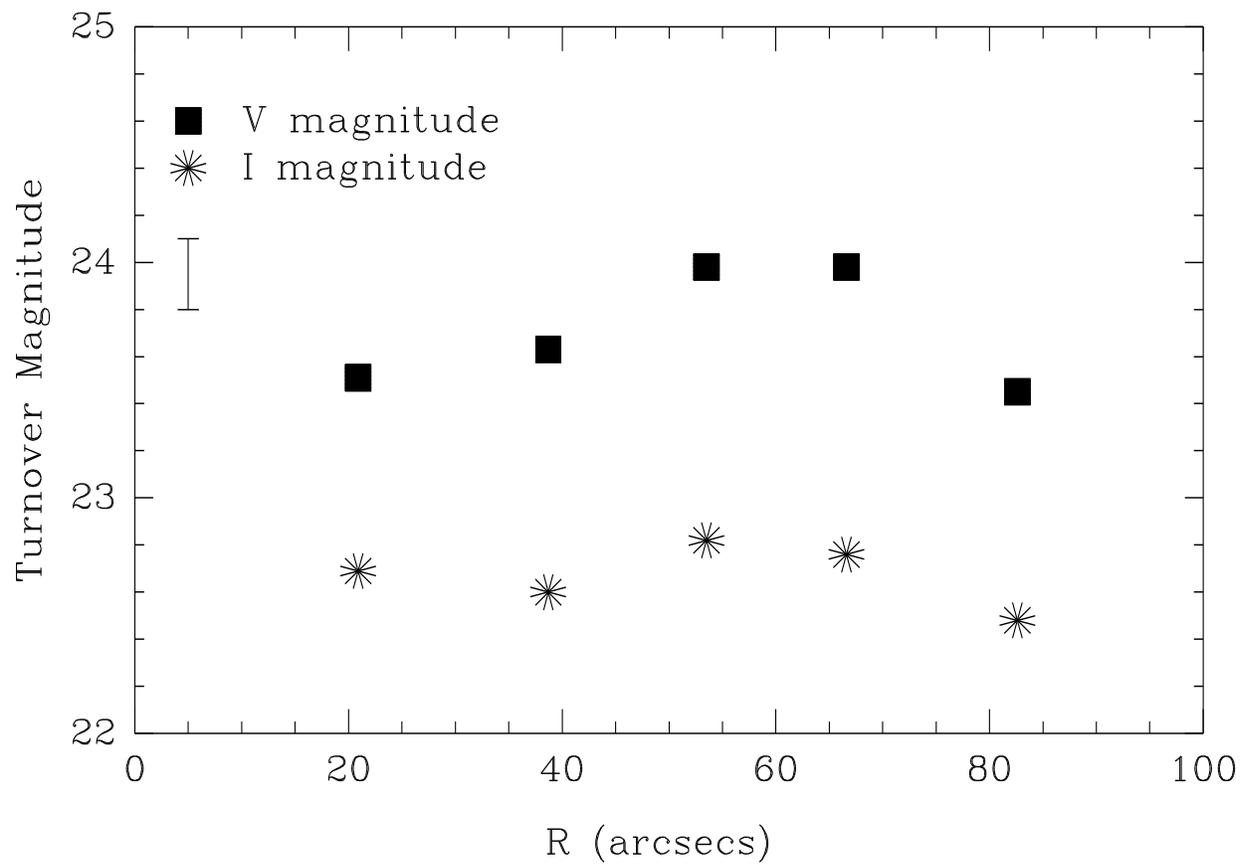
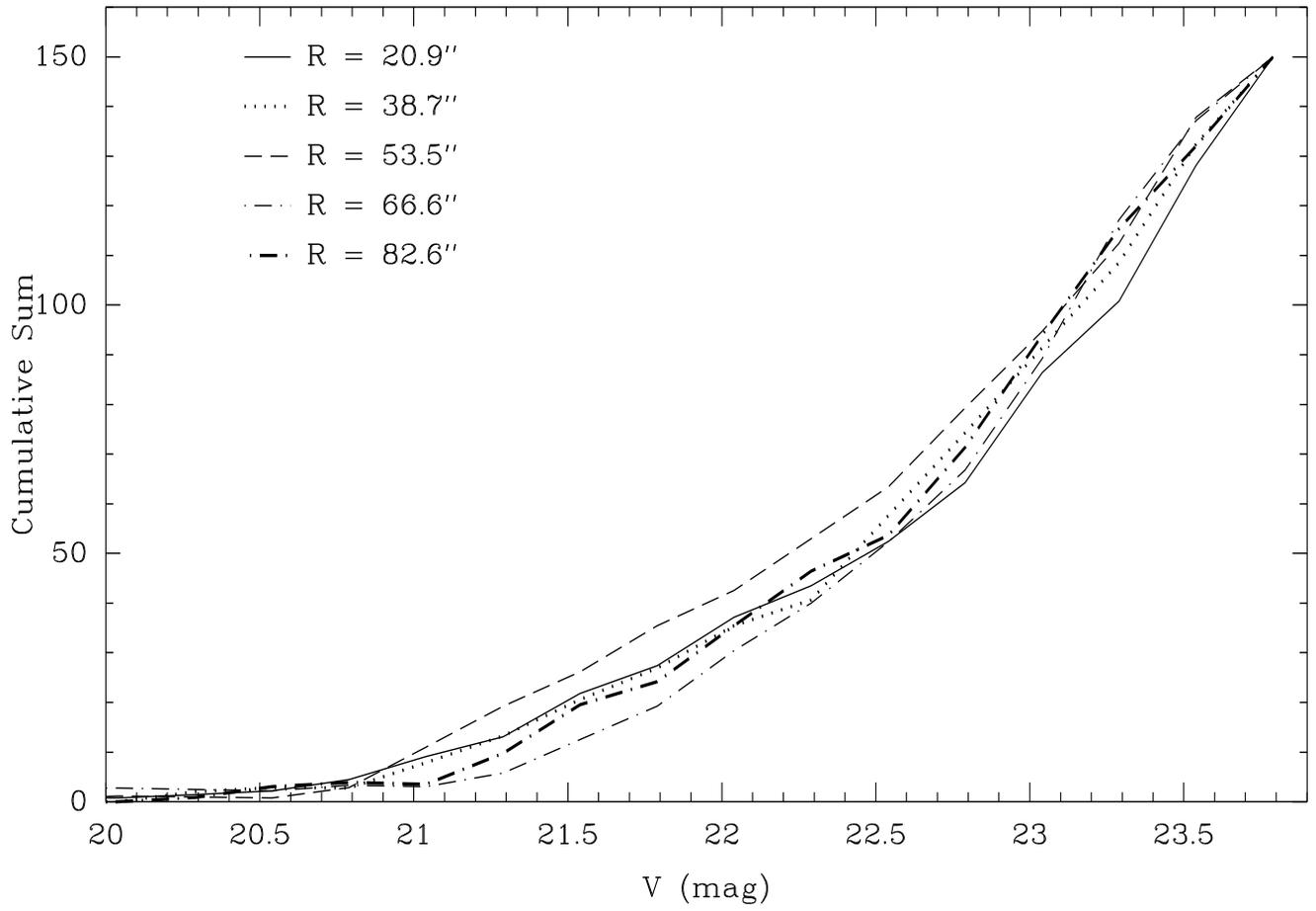
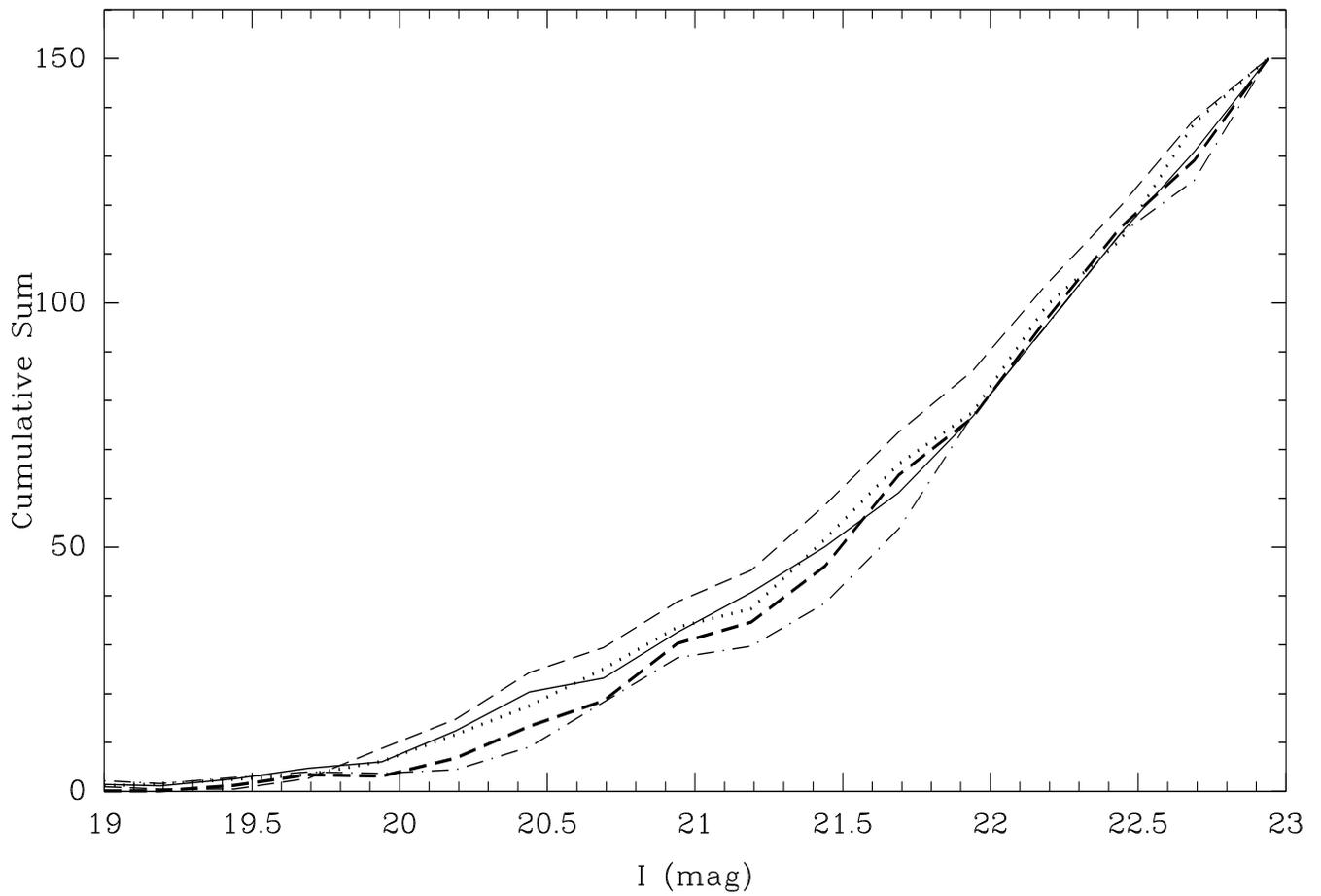
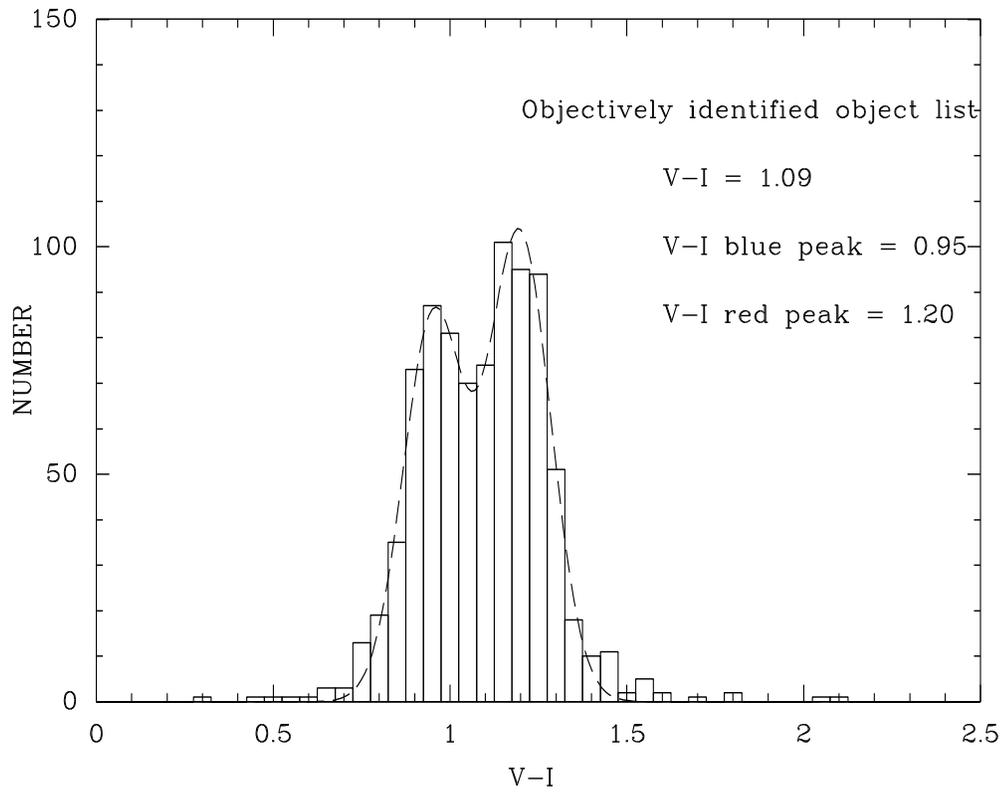
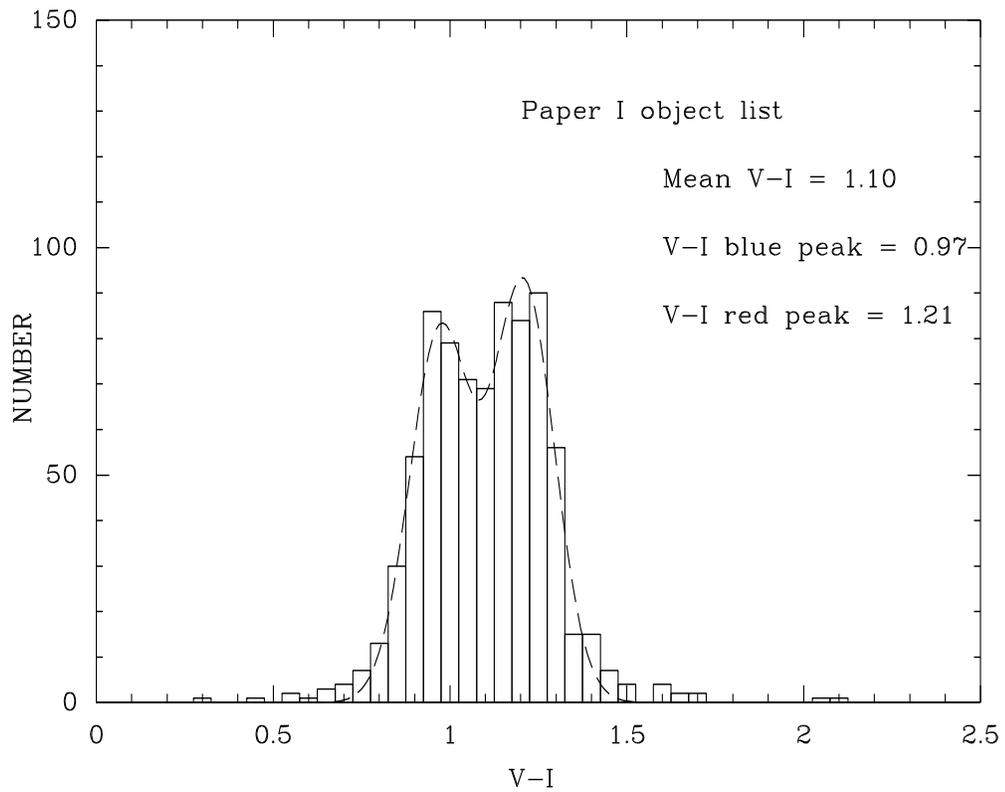
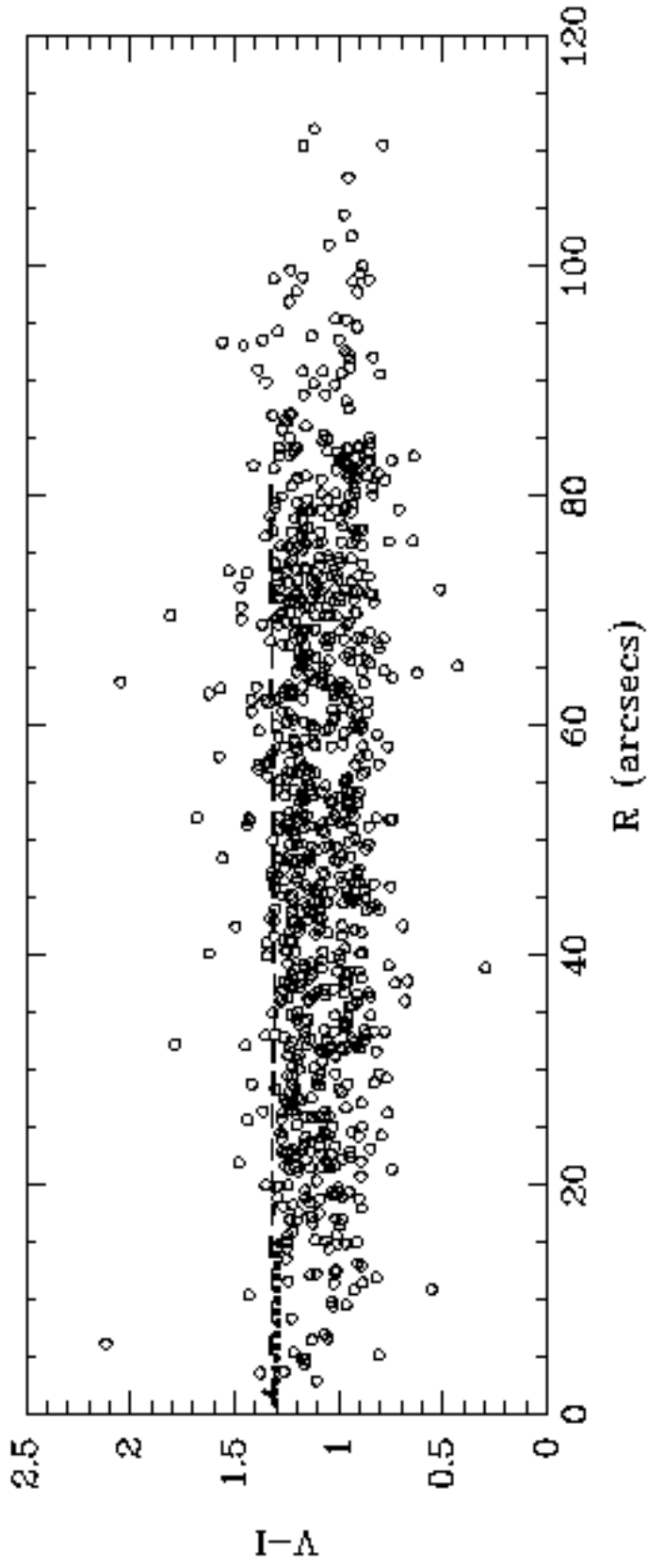
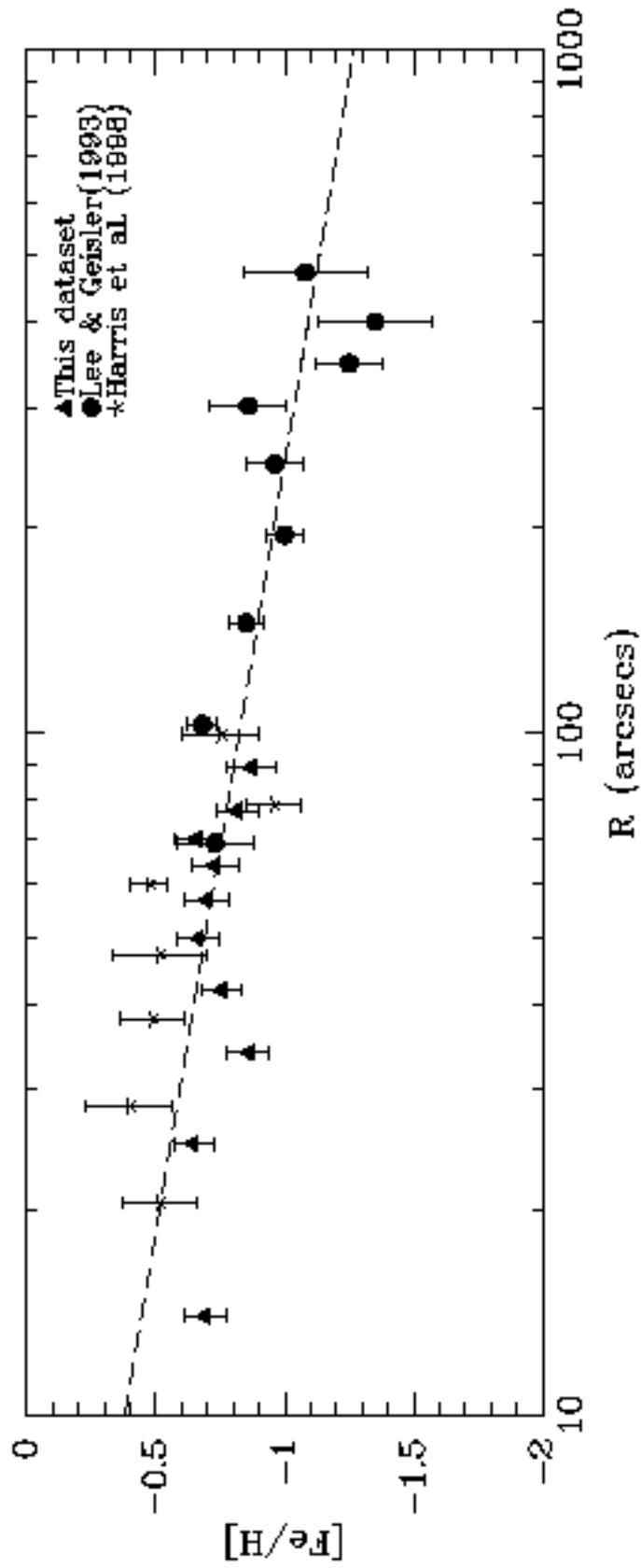
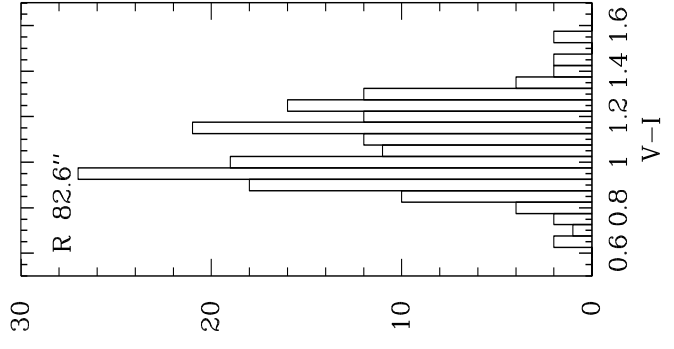
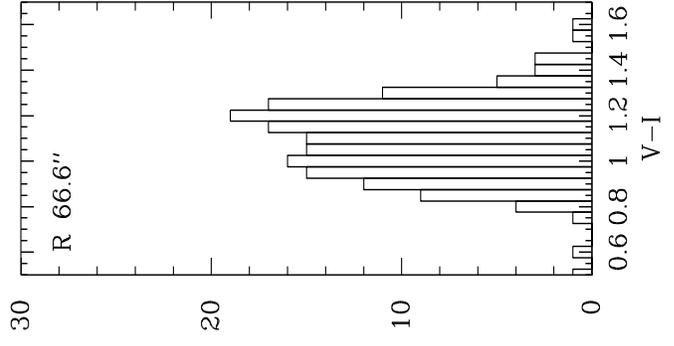
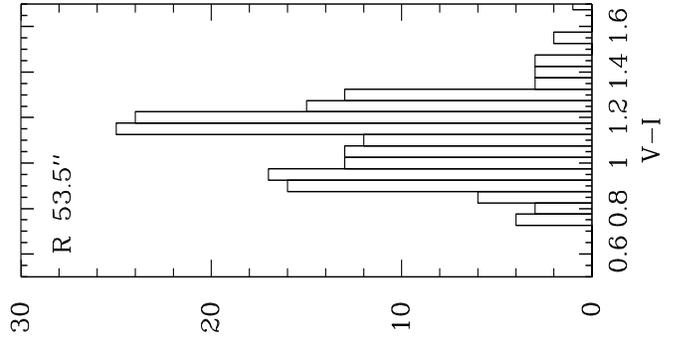
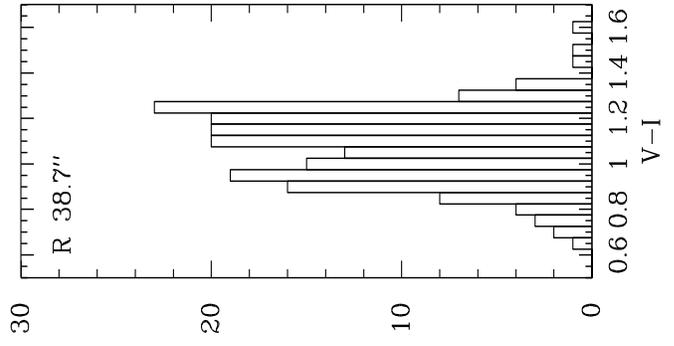
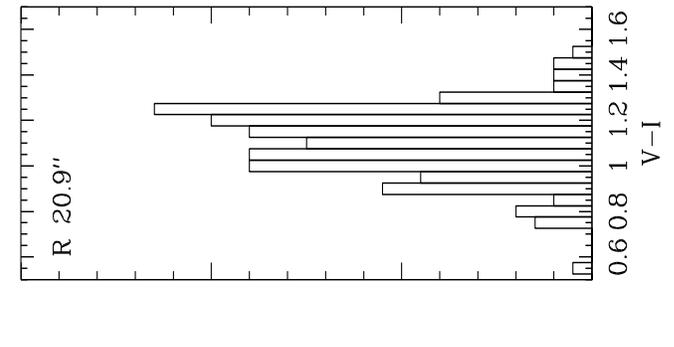
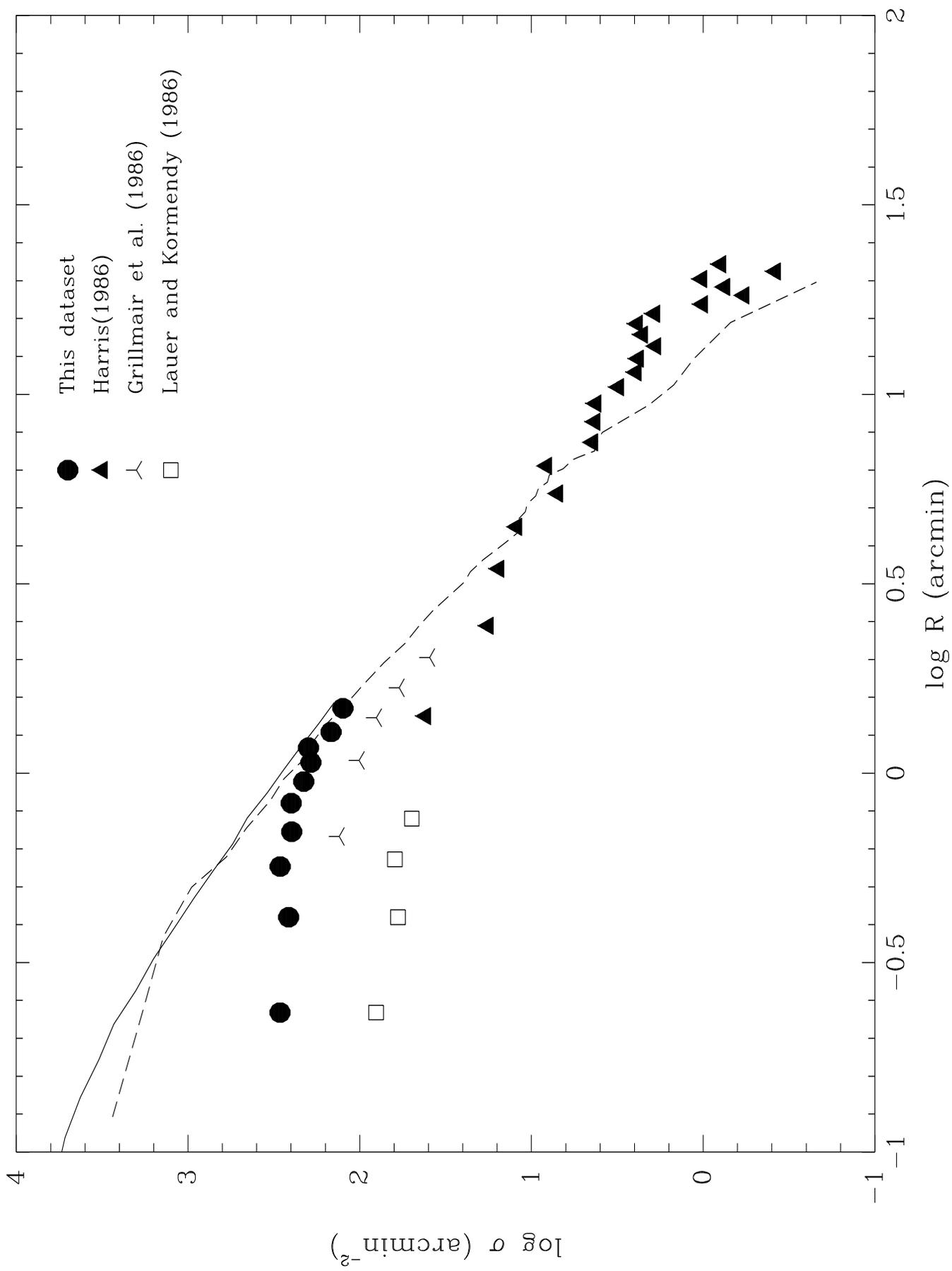
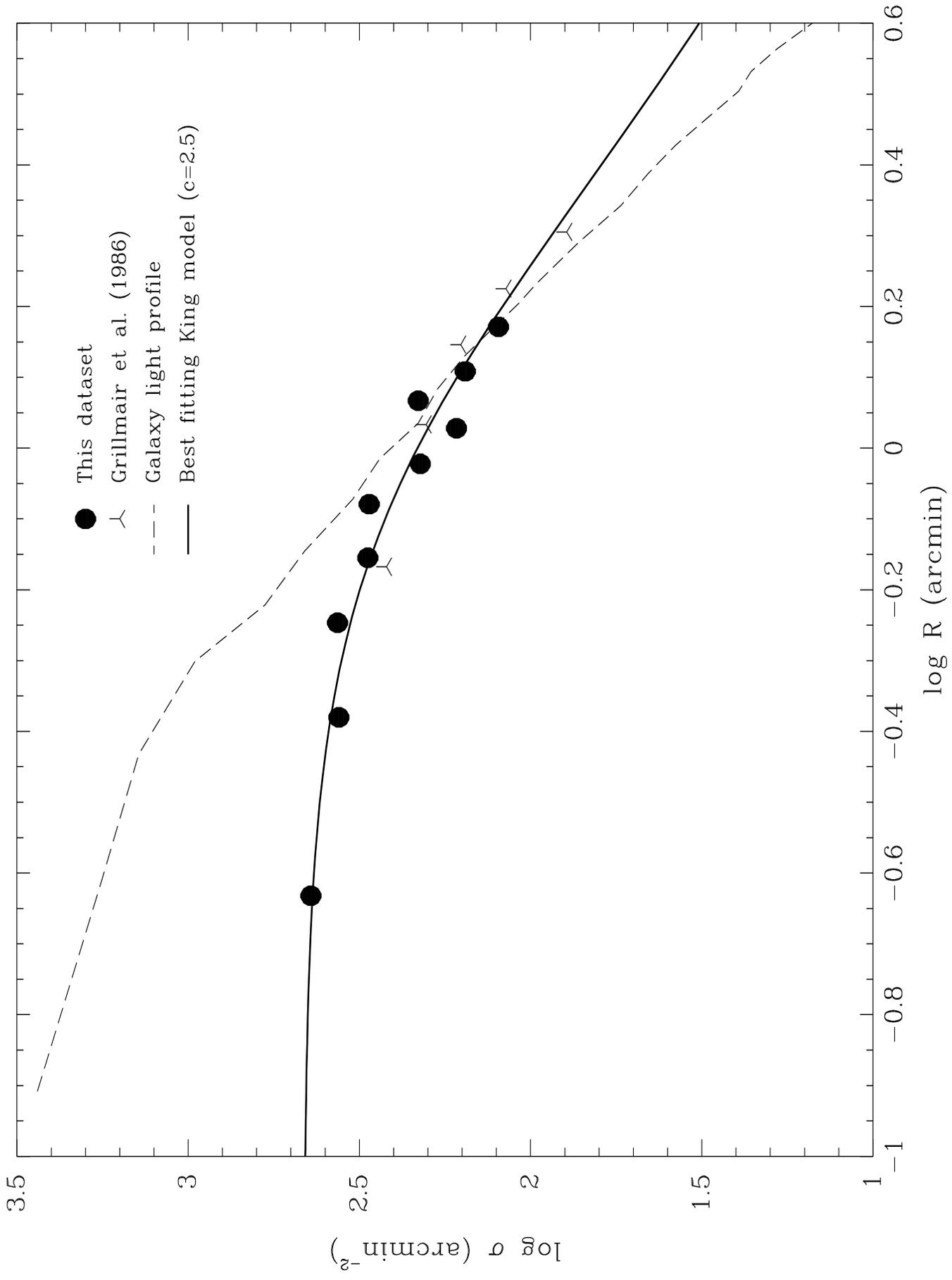
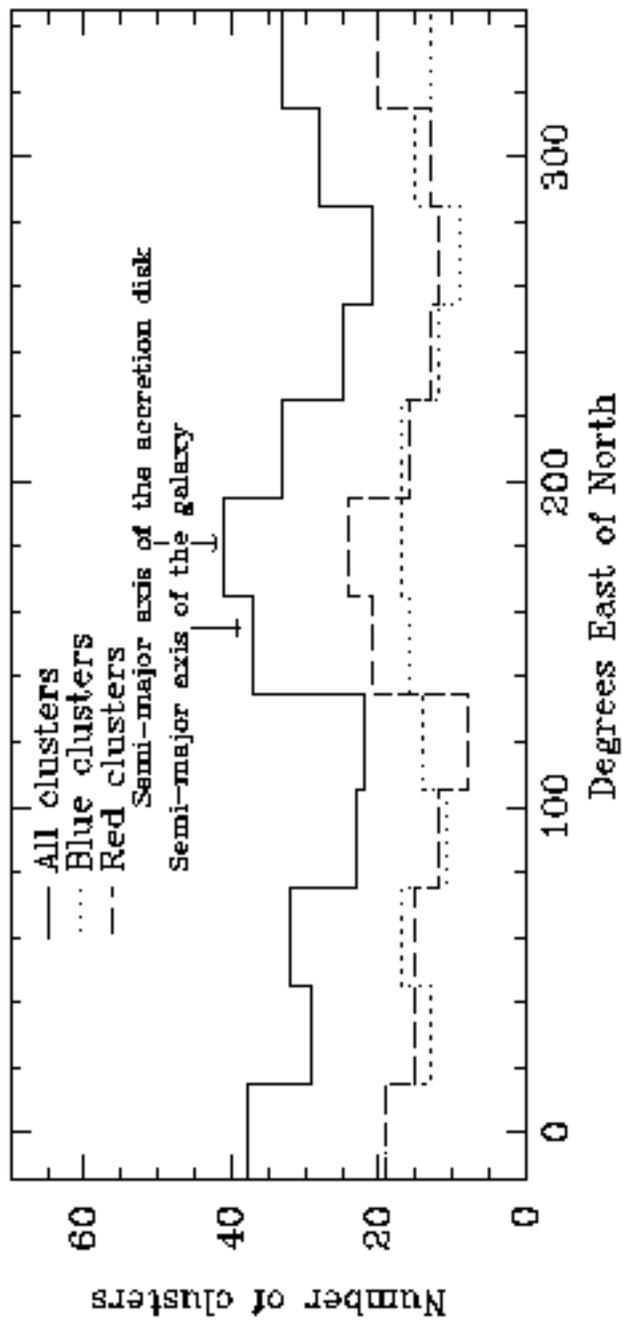
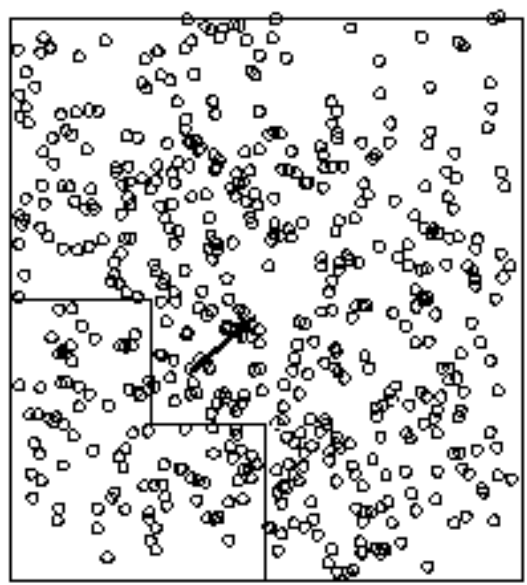
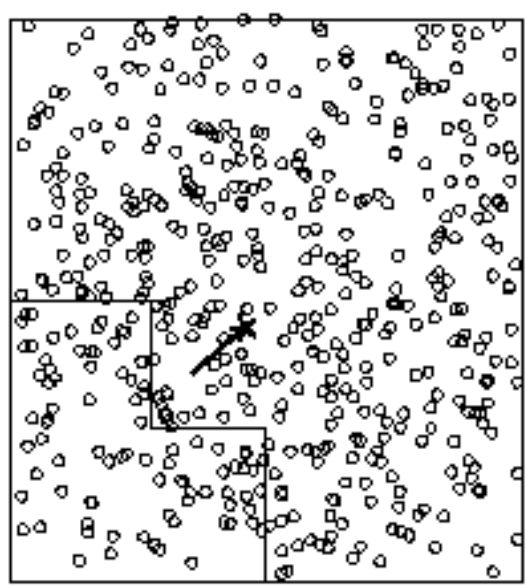
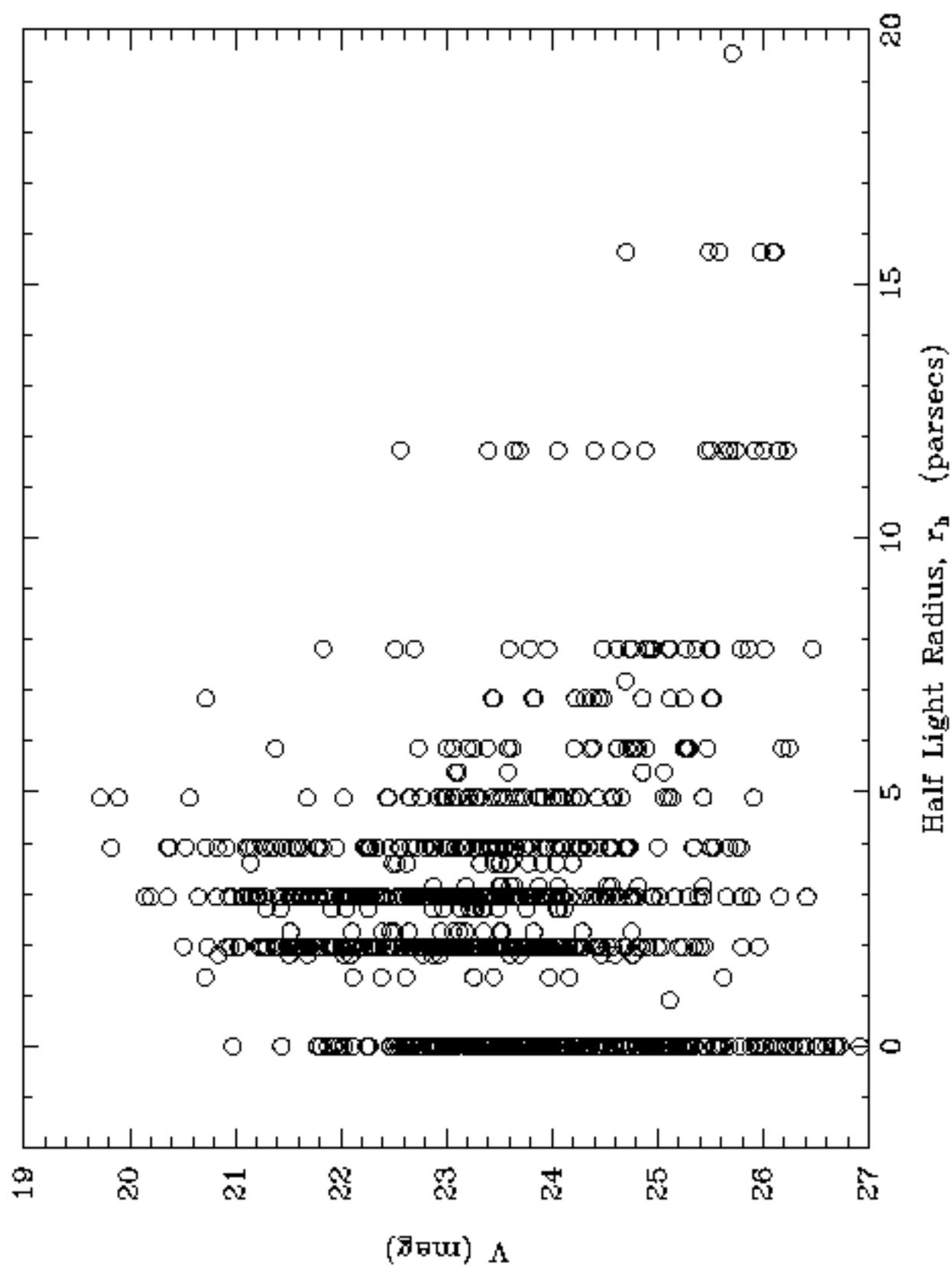
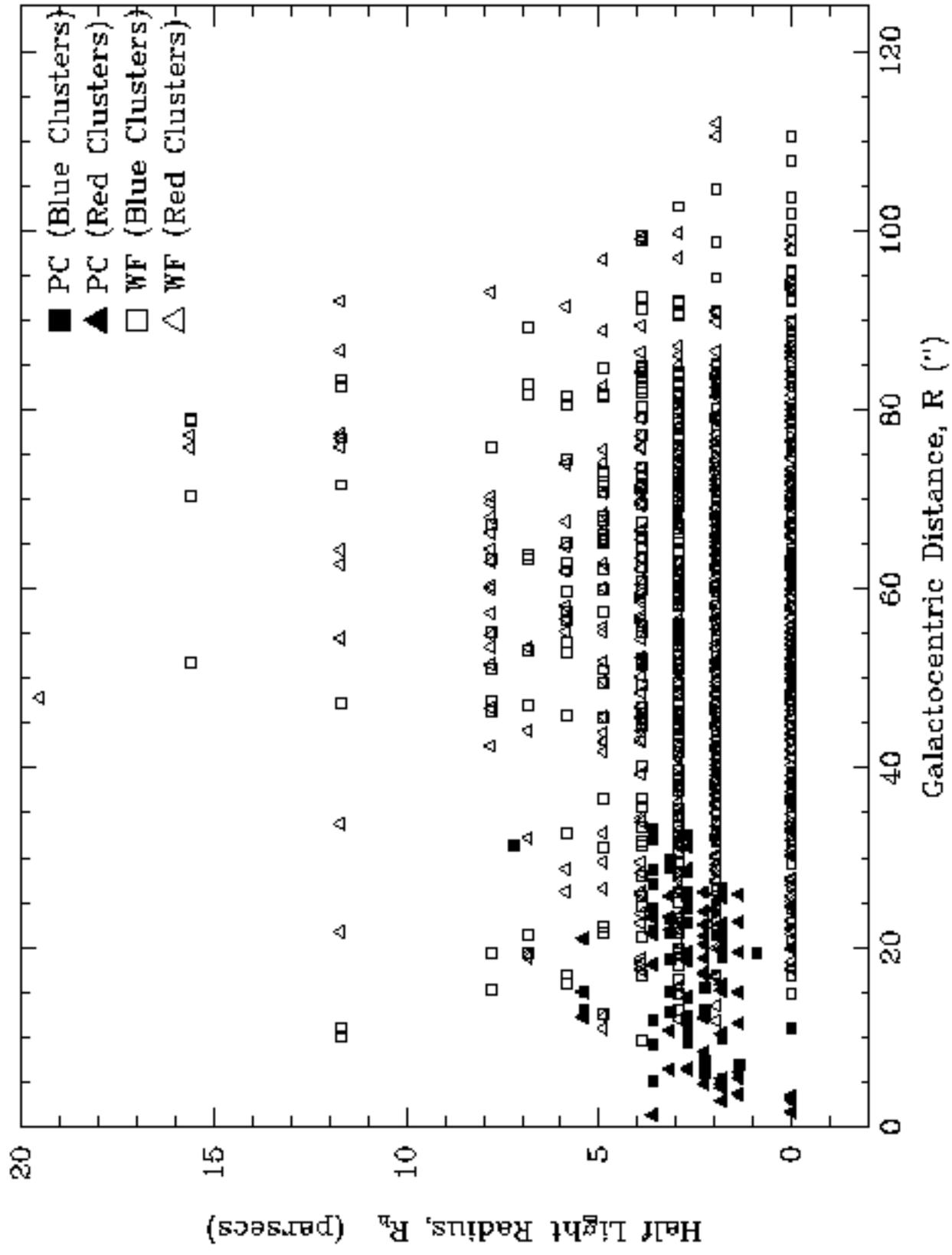
\end{document}